\documentclass[a4paper,fleqn,usenatbib,useAMS]{mnras}

\usepackage{amssymb}	
\usepackage{multicol}        
\usepackage{bm}		
\usepackage{pdflscape}	
\usepackage[T1]{fontenc}
\usepackage{ae,aecompl}
\usepackage{times}

\usepackage {graphicx}
\usepackage{natbib}
\usepackage{aas_journals}
\usepackage{color}
\usepackage{amsmath}
\newcommand{\photoz}{photo-$z$}

\newcommand{\redmagic}{\textit{redMaGiC}}
\newcommand{\ZOBOV}{\texttt{ZOBOV}}
\newcommand{\VIDE}{\texttt{VIDE}}

\def\aap{{ A\&A}}

\def\aj{{AJ}}

\def\apj{{ApJ}}
\def\apjl{{ApJL}}

\def\mnras{{MNRAS}}

\def\pasj{{PASJ}}

\def\apjs{{ApJS}}

\def\redmagic{redMaGiC}

\newcommand{\bes}{\begin{equation*}}
\newcommand{\ees}{\end{equation*}}
\newcommand{\bea}{\begin{eqnarray}}
\newcommand{\eea}{\end{eqnarray}}
\newcommand{\beas}{\begin{eqnarray*}}
\newcommand{\eeas}{\end{eqnarray*}}

\newcommand{\ltsima}{$\; \buildrel < \over \sim \;$}
\newcommand{\lsim}{\lower.5ex\hbox{\ltsima}}
\newcommand{\gtsima}{$\; \buildrel > \over \sim \;$}
\newcommand{\gsim}{\lower.5ex\hbox{\gtsima}}

\newcommand{\be}{\begin{equation}}
\newcommand{\ee}{\end{equation}}
\newcommand{\ba}{\begin{eqnarray}}
\newcommand{\ea}{\end{eqnarray}}

\title[DES Y1 voids $\times$ {\it Planck} CMB lensing]{Dark Energy Survey Year 1 Results: the lensing imprint of cosmic voids on the Cosmic Microwave Background}

\author[Vielzeuf et al.]{
\parbox{\textwidth}{
\LARGE
P.~Vielzeuf$^{1,2,3}$\thanks{Corresponding author: \texttt{\rm \texttt{pvielzeu@sissa.it}}}
A.~Kov\'{a}cs$^{1,4,5}$\thanks{Corresponding author: \texttt{\rm \texttt{akovacs@iac.es}}, Juan de la Cierva Fellow},
U.~Demirbozan$^{1}$,
P.~Fosalba$^{6,7}$,
E.~Baxter$^{8}$,
N.~Hamaus$^{9}$,
D.~Huterer$^{10}$,
R.~Miquel$^{11,1}$,
S.~Nadathur$^{12}$,
G.~Pollina$^{9}$,
C.~S{\'a}nchez$^{8}$,
L.~Whiteway$^{13}$,
T.~M.~C.~Abbott$^{14}$,
S.~Allam$^{15}$,
J.~Annis$^{15}$,
S.~Avila$^{16}$,
D.~Brooks$^{13}$,
D.~L.~Burke$^{17,18}$,
A.~Carnero~Rosell$^{19,20}$,
M.~Carrasco~Kind$^{21,22}$,
J.~Carretero$^{1}$,
R.~Cawthon$^{23}$,
M.~Costanzi$^{24,25}$,
L.~N.~da Costa$^{20,26}$,
J.~De~Vicente$^{19}$,
S.~Desai$^{27}$,
H.~T.~Diehl$^{15}$,
P.~Doel$^{13}$,
T.~F.~Eifler$^{28,29}$,
S.~Everett$^{30}$,
B.~Flaugher$^{15}$,
J.~Frieman$^{15,31}$,
J.~Garc\'ia-Bellido$^{16}$,
E.~Gaztanaga$^{6,7}$,
D.~W.~Gerdes$^{32,10}$,
D.~Gruen$^{33,17,18}$,
R.~A.~Gruendl$^{21,22}$,
J.~Gschwend$^{20,26}$,
G.~Gutierrez$^{15}$,
W.~G.~Hartley$^{13,34}$,
D.~L.~Hollowood$^{30}$,
K.~Honscheid$^{35,36}$,
D.~J.~James$^{37}$,
K.~Kuehn$^{38,39}$,
N.~Kuropatkin$^{15}$,
O.~Lahav$^{13}$,
M.~Lima$^{40,20}$,
M.~A.~G.~Maia$^{20,26}$,
M.~March$^{8}$,
J.~L.~Marshall$^{41}$,
P.~Melchior$^{42}$,
F.~Menanteau$^{21,22}$,
A.~Palmese$^{15,31}$,
F.~Paz-Chinch\'{o}n$^{21,22}$,
A.~A.~Plazas$^{42}$,
E.~Sanchez$^{19}$,
V.~Scarpine$^{15}$,
S.~Serrano$^{6,7}$,
I.~Sevilla-Noarbe$^{19}$,
M.~Smith$^{43}$,
E.~Suchyta$^{44}$,
G.~Tarle$^{10}$,
D.~Thomas$^{12}$,
J.~Weller$^{45,46,9}$,
J.~Zuntz$^{47}$
}
  \vspace{0.07cm}\\~\\
\parbox{\textwidth}{\centering \textsc{\Large(The DES Collaboration)} \\ \centering \textit{Author affiliations are listed at the end of this paper}\\ }}
\begin{document}
\pagerange{\pageref{firstpage}--\pageref{lastpage}} \pubyear{2020}
\maketitle
\label{firstpage}
\begin{abstract}
Cosmic voids gravitationally lens the cosmic microwave background (CMB) radiation, resulting in a distinct imprint on degree scales. 
We use the simulated CMB lensing convergence map from the MICE N-body simulation to calibrate our detection strategy for a given void definition and galaxy tracer density. We then identify cosmic voids in DES Year 1 data and stack the {\it Planck} 2015 lensing convergence map on their locations, probing the consistency of simulated and observed void lensing signals.
When fixing the shape of the stacked convergence profile to that calibrated from simulations, we find imprints at the $3\sigma$ significance level for various analysis choices. The best measurement strategies based on the MICE calibration process yield $S/N\approx4$ for DES Y1, and the best-fit amplitude recovered from the data is consistent with expectations from MICE ($A\approx1$). Given these results as well as the agreement between them and N-body simulations, we conclude that the previously reported excess integrated Sachs-Wolfe (ISW) signal associated with cosmic voids in DES Y1 has no counterpart in the {\it Planck} CMB lensing map.
\end{abstract}

\section{Introduction}
\label{Section1}
The standard model of cosmology is based on the assumption that our universe is homogeneous and isotropic at large scales. However, going to smaller scales one can observe a hierarchical clustering of matter that forms different structures in the cosmic web.

Surrounded by galaxies, galaxy clusters, filaments and walls, cosmic voids are large underdense regions that occupy the majority of space in our Universe. They are the most dark energy dominated regions in the cosmic web, essentially devoid of dark matter and related non-linear effects. Their underdense nature thus makes them good candidates for studying the dark energy phenomenon \citep{ryden1995,leepark2009,bos2012,pisani2015,sutter2015} and to probe its alternatives \citep{Zivick2015,cai2015,li2012,clampitt2013,verza2019}. Modified gravity models attempt to explain cosmic acceleration without the use of a cosmological constant, however the viability of these types of models imply specific screening mechanisms that will ensure the agreement with solar system observations (see e.g. \cite{braxlecture} for a review of such models). These screening mechanisms predict that the intrinsic density of high-density regions (such as dark matter halos) and low-density regions (such as cosmic voids as the unscreened regime) will be more and less dense, respectively, than in the general relativity scenario \citep{Martinosheth2009}. Similarly, measuring the underlying matter profile of these structures appears to be an interesting tool to probe cosmological models \citep{Cautun2018}.

The lensing signals by individual voids are difficult to detect \cite[see e.g.][]{Amendola1999}, but recent work has shown that a stacking methodology could help to increase the signal-to-noise ratio and make the detection possible \citep{Krause2013,Davies2018,Higuchi2013}. Recently, the void lensing signal has been observed in galaxy shear statistics \citep{melchior2014,clampittjain2015,Carles_void} with moderate significance ($\sim 4.4-7\sigma$). The most significant detection to date is $14\sigma$ by \cite{Fang2019} using the Dark Energy Survey first year data set \citep[DES Y1][]{DES}. Similarly, a lensing signal has been detected using projected underdense regions, the so-called troughs \citep{troughsSV,troughKIDS} with higher significance ($\sim 10-15 \sigma$ detection). 
The above studies used shear statistics of galaxies around void centres, i.e. measured anisotropic shape deformations of galaxies due the gravitational field inside voids. Likewise, these weak lensing imprints are also expected to be observed in the reconstructed lensing convergence maps of the cosmic microwave background radiation (CMB hereafter). In particular, cosmic voids cause a de-magnification effect and therefore correspond to local minima in the convergence maps.

The CMB lensing imprint of other elements of the cosmic web have also been measured recently. \cite{Madhavacheril2015} detected the lensing of the CMB by optically-selected galaxies. Along similar lines, \cite{Baxter2015} detected a CMB lensing effect by galaxy clusters selected from the South Pole Telescope (SPT) data, while \cite{He2017} reconstructed the correlation of filamentary structures in the Sloan Digital Sky Survey (SDSS) data and CMB lensing convergence ($\kappa$, hereafter), as seen by {\it Planck}. Then, \cite{Baxter2018} stacked the SPT $\kappa$ maps on locations of galaxy clusters identified in DES Y1 data, finding good consistency between simulated and observed results.

The prospects of obtaining cosmological parameter constraints from CMB lensing probed using cosmic voids are discussed by \cite{Chantavat2016,chantavat2017}. The role of void definition, environment, and type have also been studied in simulations. \cite{nadathur2017} found that the sensitivity of the detection of void lensing effects could be significantly improved by considering sub-populations.

Following their own stacking measurement strategy, \cite{cai} have, for the first time, detected a CMB lensing signal using cosmic voids \citep[catalogue created by][]{Mao2017} identified in the CMASS (constant mass) galaxy tracer catalogue of the Baryon Oscillation Spectroscopic Survey Data Release 12 (BOSS DR12). They identified voids using the {\ZOBOV} (ZOnes Bordering On Voidness) void finder algorithm \citep{zobov}. Their main aim was to complement the stacking measurements of the integrated Sachs-Wolfe effect (ISW) \citep{SachsWolfe} using the CMB lensing analyses. Evidences for ISW and CMB lensing imprints of the same cosmic voids helps to confirm the reality of each effect. They discuss that an ISW-lensing dual probe is valuable from the point of view of modified gravity, since the two effects are closely related: lensing depends on the sum of metric potentials, whereas ISW depends on their time derivatives. Hints of the correlations between the ISW and CMB lensing signatures have also been found by the {\it Planck} team \citep{PlanckISW2015}. Nevertheless, \cite{cai} reported a CMB lensing signal of BOSS voids that is compatible with simulated imprints, with somewhat higher-than-expected signal in the centre of the voids. The conclusion was that the puzzling excess ISW signal was seen in the BOSS DR12 data, especially for the most significant large and deep voids, but the lensing counterpart seemed inconclusively noisy with hints of a small excess imprint.

In the first year footprint of the Dark Energy Survey, \cite{Kovacs2016} have recently attempted to probe these claims in a different part of the sky and identified 52 voids and 102 superclusters at redshifts $0.2 < z < 0.65$ using the void finder tool described in \cite{Carles_void}. 

These measurements, although hinting again at a large ISW amplitude, were indecisive because of the significant noise level in the stacked images of the CMB. More recently, \cite{Kovacs2019} extended the Year-1 ISW stacking analysis to Year-3 DES data, and confirmed the ISW excess signal of supervoids with higher confidence. In this paper, we aim to probe the detection of the excess ISW signals in the DES Y1 data from another point of view by measuring the corresponding CMB lensing imprint of the voids.

The paper is organised as follows. In Section \ref{Section1} we motivate our analysis of DES voids and discuss the relevance of the problem. Section \ref{Section2} is dedicated to the description of the simulation and data products that we use for our cross-correlations. In Section \ref{sec:voids_finding}, we explain how cosmic voids are identified in our study and we discuss details about the resulting void catalogues and properties of voids. Then, in Section \ref{Section4} we detail our actual cross-correlation measurement method using cosmic voids and lensing convergence maps. We present our original results in Section \ref{Section5} with details in observational cross-correlation results and their consistency tests with respect to simulated DES data. Finally, Section \ref{Section6} contains a summary and discussion of our most important findings and their significance, including a vision of possible future projects.

\section{Data sets and simulations}\label{Section2}

In this section, we introduce the galaxy tracer catalogues and corresponding lensing convergence maps that we aim to cross-correlate.

\subsection{Observations - DES Y1 {\redmagic} catalogues}

We use photometric redshift data from the Dark Energy Survey (DES) to identify cosmic voids. DES is a six-year survey. After the first three years of data (Y3)
DES covers about one eighth of the sky (5000 deg$^2$) to a depth of $i_{AB} < 24$, imaging about 300 million galaxies in 5 broadband filters ($grizY$) up to redshift $z=1.4$ \citep[for details see e.g.][]{DECam,morethanDE2016}.

In this paper we used a luminous red galaxy sample from the first year of observations (Y1, 1300 deg$^2$ survey area). This Red-sequence MAtched-filter Galaxy Catalogue \cite[\redmagic, see][for DES science verification (SV) test results]{Rozo2015} is a catalogue of photometrically selected luminous red galaxies, based on the red-sequence MAtched-filter Probabilistic Percolation (redMaPPer) cluster finder algorithm \citep{Rykoff2014}. See also \cite{wthetapaper} for further details about the DES Y1 {\redmagic} sample that is not identical to the SV data set.

An important source of error that affects our void finding procedure is {\photoz} uncertainty \citep[see e.g.][and references therein]{Hoyle2018}; photometric DES data does not provide a precise redshift estimate for the galaxy tracers of voids in the way that a spectroscopic survey does \citep[see e.g.][]{Carles_void}. However, the \redmagic\ sample has exquisite photometric redshifts, namely $\sigma_z/(1+z)\approx 0.02$, and a $4\sigma$ redshift outlier rate of $r_\mathrm{out}\simeq1.41\%$. The resulting galaxy sample has a constant co-moving space density in three versions, $\bar{n}\approx 10^{-3}h^{3}$ Mpc$^{-3}$ (high density sample, brighter than 0.5$L_{*}$), $\bar{n}\approx4\times10^{-4}h^{3}$ Mpc$^{-3}$ (high luminosity sample, brighter than 1.0$L_{*}$)), $\bar{n}\approx1\times10^{-4}h^{3}$ Mpc$^{-3}$ (higher luminosity sample, brighter than 1.5$L_{*}$)). For further details about the \redmagic\ sample see \cite{Rozo2015} and \cite{wthetapaper}.

We will argue in section \ref{sec:voids_finding} that significant real underdensities can be identified even using photo-$z$ data.

\subsection{Observations -  \emph{Planck} CMB lensing}\label{sec:planck_conv}

Similarly to the distortion of galaxy shapes due to the underlying matter field, used in weak lensing analyses, CMB photons also experience deflections along their path and thus will be observed lensed. In general, one can express the lensed (observed) CMB temperature $T$ in the $\hat{n}$ direction as a deviation (remapping) of the un-lensed (emitted) temperature $\Tilde{T}$:
\begin{equation}
    T(\hat{n})=\Tilde{T}(\hat{n}+\Vec{\alpha}(\hat{n})),
\end{equation}
where $\Vec{\alpha}(\hat{n})$ is the deflection angle that is used to define a lensing potential via $\alpha=\nabla\Phi(\hat{n})$  \citep[see e.g.][for details]{lewischallinor2006}. Assuming a flat universe, the CMB lensing potential in a direction $\hat{n}$ is defined as:
\begin{equation}
    \Phi(\hat{n})=-2\int_0^{\chi_{\rm cmb}}d\chi\frac{\chi_{\rm cmb}-\chi}{\chi_{\rm cmb}\chi}\Psi({\chi\ \hat n;t}),
\end{equation}
where $\chi$ is the co-moving distance ($\chi_{\rm cmb}$ is the co-moving distance to the CMB) and $\Psi$ is the gravitational potential evaluated in the $\hat n$ direction and at time $t=\eta_0-\eta$ where $\eta_0$ is the conformal time today.

The gravitational potential can then be expressed as function of the underlying matter density ($\delta({\chi\hat n;t})$) field through the Poisson equation :
\begin{equation}
    \nabla^2\Psi({\chi\hat n;t})=\frac{3H_0^2\Omega_m}{2a(\eta)}\delta({\chi \hat n;t}),
\end{equation}
where $H_0$ is the expansion rate today, $\Omega_m$ is the matter energy density, and $a(\eta)$ is the scale factor evaluated at the conformal time $\eta$ (assuming here natural units, i.e. where $c=1$).

The lensing convergence, as a main observable, is defined as $\kappa=\nabla^2\Phi$, which in harmonic space can be related to the lensing potential as
\begin{equation}
\kappa_{LM}=-\frac{L}{2}L(L+1)\Phi_{LM}
\end{equation}
where $L$ and $M$ are indices of spherical harmonics of the reconstructed lensing maps. The {\it Planck} collaboration released the $\kappa_{\rm LM}$ coefficients \citep[see][for details]{Plancklensingmap} they reconstructed from their data up to  $L_{\rm max}=2048$\footnote{https://wiki.cosmos.esa.int/planck-legacy-archive/index.php/Lensing}. A convergence map can then be created by converting the $\kappa_{\rm LM}$ values into \texttt{healpix} maps \citep{Healpix}. We thus created a $\kappa$ map at $N_{\rm side}=512$ resolution using the {\it Planck} 2015 data products following \cite{cai}\footnote{We exchanged simple computer scripts with Cai et al. in order to reproduce the $\kappa$ map they used in their similar analysis, making comparisons easier. The lensing reconstruction in the {\it Planck} 2015 data release was based on minimum variance (MV) methods.}. We also constructed a corresponding mask from the publicly available {\it Planck} 2015 lensing products. We note that even though higher resolution maps may be extracted from the $\kappa_{\rm LM}$ coefficients, $N_{\rm side}=512$ is a sufficient choice given the degree-size angular scales involved in our problem.

\subsection{Simulations - the MICE galaxy mock and $\kappa$ map}
\label{mice_sims}
The MICE (Marenostrum Institut de Ciencias de l'Espai) simulated sample is an N-body light-cone extracted from the {\it MICE Grand Challenge} (MICE-GC), that contains about 70 billion dark-matter particles in a $(3 h^{-1}Gpc)^3$ comoving volume. MICE was developed at the Marenostrum supercomputer at the Barcelona Supercomputing Center (BSC)\footnote{{\bf \url{www.bsc.es.}}} running the \texttt{GADGET2} \citep{springel2005} code. The MICE mock galaxy catalogue was created and validated to follow local observational constraints such as luminosity functions, galaxy clustering (with respect to different galaxy populations), and color-magnitude diagrams. For details on the creation of the MICE simulation see e.g. \cite{fosalba2015a,crocce2015,fosalba2015}.

In particular, the {\redmagic} algorithm was run on the MICE mock galaxy catalogue following the methodology applied to observed DES Y1 data. We utilized this MICE-{\redmagic} mock galaxy catalogue to trace the large-scale galaxy distribution and to identify cosmic voids. Positions of cosmic voids in MICE were then cross-correlated with lensing maps of the MICE simulation, produced using the ``Onion Universe'' methodology \cite[see][]{fosalba2008}. An all-sky map was constructed by cloning the simulation box from the MICE simulation. The box was translated around the observer, then the light cone was split into concentric shells in redshift bins of $\Delta z \sim 0.003 (1+z)$ size and an angular resolution of $\Delta\theta\sim0.85\arcmin$. The resulting lensing map was then validated using auto- and cross-correlations with foreground MICE galaxy and dark matter particles (see \cite{fosalba2015} for details).

The MICE $\kappa$ map was provided with a pixel resolution of $N_{\rm side}=2048$. However, given the nature of our problem and the large angular size of voids, we downgraded the high resolution map to a lower $N_{\rm side}=512$ resolution. The downgraded map matches the resolution of the {\it Planck} $\kappa$ map that we used in this analysis.

The simulation assumed a flat standard $\Lambda CDM$ model with input fiducial parameters $\Omega_m=0.25$, $\Omega_\Lambda=0.75$, $\Omega_b=0.044$, $n_s=0.95$, $\sigma_8=0.8$ and $h=0.7$ from the Five-Year Wilkinson Microwave Anisotropy Probe (WMAP) best fit results \citep{WMAPparam2009}. 

We note that the MICE cosmology is relatively far from the best-fit {\it Planck} cosmology \citep{Planck2018_cosmo}. For instance, they differ in the values of $\Omega_m$ and the Hubble constant $H_0$ that are expected to affect the amplitude of the lensing signal (see Eq. 3). Possibly, these parameter differences also affect the shape of the imprint profile if voids themselves evolve significantly differently in different cosmologies.

In this study, we assume that these differences in cosmological parameters are negligible in our analysis but nevertheless perform additional tests using the WebSky simulation \citep{websky}. This simulation package provides mock light-cone catalogues and corresponding CMB lensing convergence information assuming the best-fit {\it Planck} 2018 cosmology (for further details about the MICE-WebSky comparison see Section \ref{websky_tests}).

We note that the most influential parameter for determining the matter content and the lensing convergence of voids in fact appears to be $\sigma_8$ \citep[see][for details]{Nadathur2019}. Relevantly, its value in the MICE simulation ($\sigma_8=0.8$) is quite close to the best-fit {\it Planck} value ($\sigma_8=0.811\pm0.006$).
 
On the other hand, cosmological constraints from modern weak lensing surveys such as KiDS (Kilo-Degree Survey) \citep[see e.g.][]{Hildebrandt2020,Joudaki2019}, the Subaru Hyper Suprime-Cam survey \citep{HSC}, or the DES Y1 data itself \citep{DESY1_cosmo} prefer lower matter densities than the best-fit {\it Planck} cosmology (with $\Omega_m\approx0.26\pm0.03$ in the case of DES Y1). In particular, these findings demand lower values for the $S_{8}=\sigma_{8}\sqrt{\Omega_{m}/0.3}$ parameter that are discordant with the corresponding {\it Planck} value at the $2.5\sigma$ level. Note that these lower matter densities are consistent with the MICE value of $\Omega_{m}=0.25$, suggesting that it represents a good description of the DES Y1 data that we wish to use in this analysis.

As an example, we mention that, in a similar DES analysis, \cite{Fang2019} reported good qualitative agreement between the weak lensing signal of simulated MICE voids and observed DES Y1 voids, in spite of these potential differences.

\section{Void catalogues}\label{sec:voids_finding}

The identification of voids in the cosmic web is an intricate process that is affected by specific survey properties such as tracer quality, tracer density, and masking effects \citep{sutter2012,Sutter_bias,sutter2014,pollina2017,Pollina2018}. The void properties also depend significantly on the methodology used to define the voids \citep[see e.g.][]{Nadathur2015}. For instance, \cite{finder_comparison} compared several void finders in a simulation using different tracers (dark matter halos, galaxies, clusters of galaxies). More recently, \cite{Cautun2018} studied how modified gravity models can be tested with different void definitions. They concluded that void lensing observables are better indicators for tests of gravity if defined in 2D projection, where the projected voids lead to ``tunnels" or ``troughs". Such definitions are particularly promising in {\photoz} surveys, such as DES, given the significant smearing effect of redshift uncertainties. However, recent advances on applying 3D void finders to DES data have demonstrated their ability to identify orders of magnitude more voids and therefore to allow precise lensing measurements \citep[see][for details]{Fang2019}.

\subsection{A 2D void finder optimized for {\photoz} data}

\cite{Carles_void} showed that significant real underdensities can be identified even using photo-$z$ data in tomographic slices of width roughly twice the typical photo-$z$ uncertainty \citep[see also][for DES tests proving that even 3D voids can be identified in photo-$z$ data in cluster tracer samples]{Pollina2018}. The heart of the method by \cite{Carles_void} is a restriction to 2D slices of galaxy data, and measurements of the projected density field around centres defined by minima in the corresponding smoothed density field. The line-of-sight slicing was found to be appropriate for slices of thickness $2s_{v}\approx100~{\rm Mpc/h}$ for photo-$z$ errors at the level of $\sigma_{z}/(1+z) \approx 0.02$ or $\sim50~{\rm Mpc/h}$ at $z\approx0.5$. 

A free parameter in the method is the scale of the initial Gaussian smoothing applied to the projected galaxy density field. A direct consequence of a larger smoothing is the merging of neighbour voids into more extended but shallower underdense structures (supervoids) where possible. The merging properties of voids, therefore, encode information about the gravitational potential of their environment. In turn, details of the potential in and around voids affect void observables including lensing imprints \citep{nadathur2017}.
For instance, \cite{Kovacs2016} found that $\sigma=20~{\rm Mpc/h}$ is a preferable choice for ISW measurements using the whole void sample in the stacking procedure. For weak galaxy lensing measurements with DES voids, however, \cite{Carles_void} reported that the smaller $\sigma=10~{\rm Mpc/h}$ smoothing is preferable. We will test how the choice of the smoothing parameter affects the void catalogue properties themselves and their CMB lensing signals.

We have also applied the random point methodology of \cite{clampittjain2015} to eliminate voids in the edges with a potential risk of mask effects. This method evaluates the density of random points that have been drawn within the mask inside each void. Then, voids with a significant part of their volume laying outside the mask will have a lower random point density and therefore can be identified and excluded. In the final catalogue, we have also excluded voids of radius $r_{v}<20~{\rm Mpc/h}$ that are expected to be spurious given the photometric redshift uncertainties.

We determine the radius of each void as well as its redshift (defined as the mean redshift of the slice in which it has been identified). The void finder operates and then defines the $r_{v}$ void radius as follows:
\begin{enumerate}
\item Divides the sample in redshift slices of size $s$ in co-moving distances. As in \cite{Carles_void}, we choose slices of size 100 Mpc/h that have shown to be a good compromise between resolution in the void's line of sight position and the {\photoz} scatter. Moreover, one can also build a sample placing the slices at different positions.
\item Computes the density field for each slice by counting the number of galaxies in each pixel and smoothing the field with a Gaussian filter with a predefined smoothing scale.
\item Selects the most underdense pixel and measures the average density in concentric annuli around it. The void radius $r_{v}$ will then be defined where the mean density is reached.
\item Finally the algorithm saves the void and erases its pixels from the density slice. Then it reiterates the process with the following underdense pixel.
\end{enumerate}

The void finder also provides two additional characteristic quantities related to the under-density of voids:
\begin{itemize}
    \item {\it the mean density contrast:} $\bar{\delta} (r<r_v) = \rho/\bar{\rho}-1$ where $\rho$ is the mean density inside the void and $\bar{\rho}$ is the mean density of the corresponding redshift slice;
    \item {\it the central density contrast:} The density contrast evaluated at one quarter of the void radius $\delta_{\rm 1/4}=\delta(r=0.25r_v)$. 
\end{itemize}

\begin{figure*}
\begin{center}
\includegraphics[width=180mm]{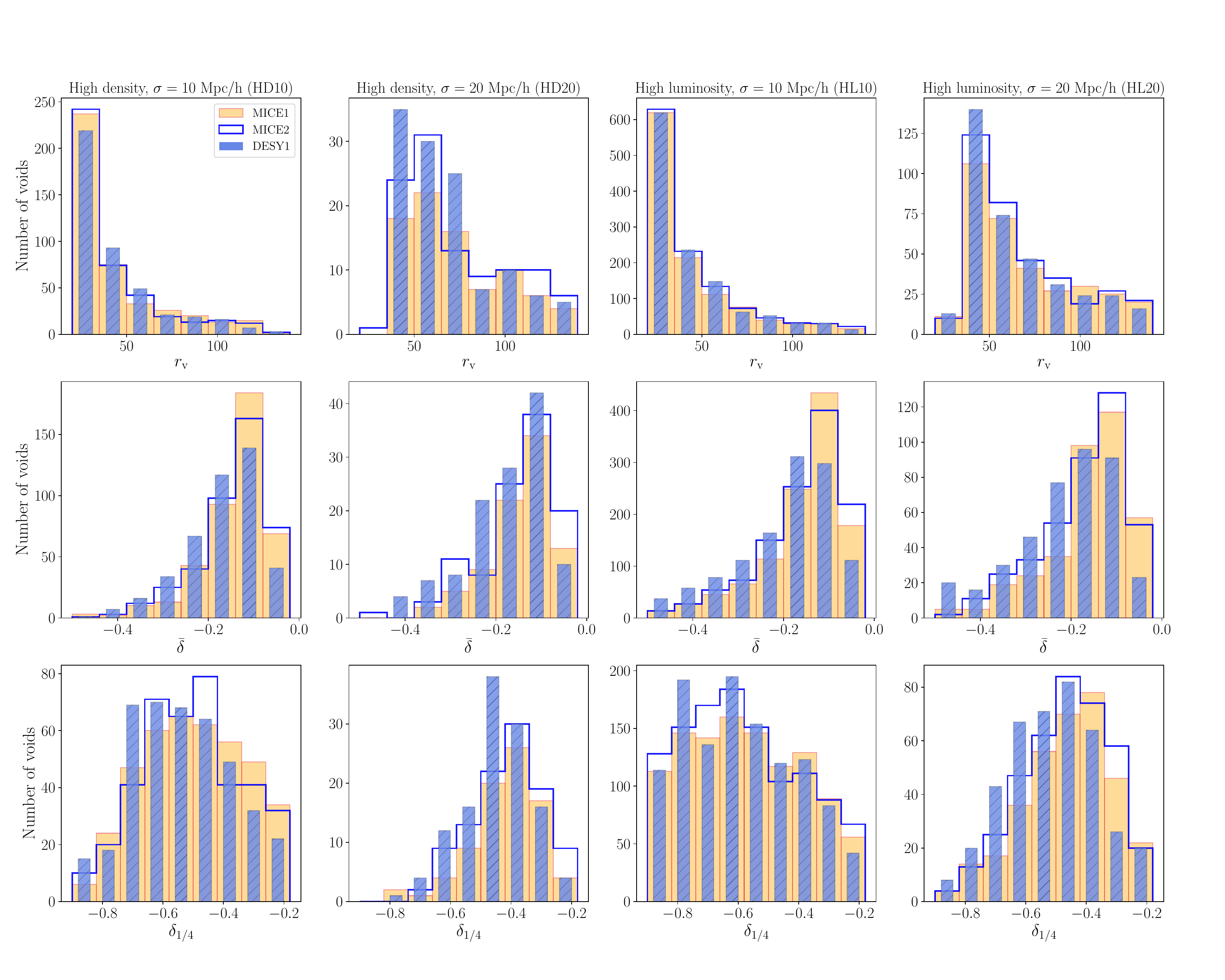}
\end{center}
\caption{Comparison of the 2D void catalogue characteristics constructed in simulated MICE1 and MICE2 (orange bars and blue steps) and observed DES Y1 samples (blue bars) with the different void catalogue versions (HD10, HD20, HL10, HL20). We present results for the high-density sample (first and second columns) and the high-luminosity sample (third and fourth columns) for different void finder smoothing scales of $10$ Mpc/h and $20$ Mpc/h.}
\label{fig:SIM_DATA_void_compa}
\end{figure*}

\subsection{{\VIDE} voids in 3-dimensional photo-$z$ data}

For completeness, we consider a different definition of voids to probe the effect on the CMB lensing signal. Namely, we employ a watershed-based finding procedure firstly introduced by the ZOnes Bordering On Voidness ({\ZOBOV}) algorithm. {\ZOBOV} defines voids using the 3-dimensional positions of the tracers \citep{zobov}. This method was originally developed to identify voids in simulations with periodic boundary conditions. However we use a {\ZOBOV} wrapper and enhanced version, the Void IDentification and Examination toolkit ({\VIDE}, hereafter) by \cite{sutter2015}, that also allows to find voids in observed data. We detail the algorithm below:
\begin{itemize}
 \item{First, the 3D tracer distribution is used as an input to construct a Voronoi tessellation. This procedure defines a volume for each particle \textit{p} consisting of all points that are closer to \textit{p} than to any other particle. In this context, the density of each cell is inversely related to its size.}

 \item{Local density minima are found by comparing cell sizes. A Voronoi cell that is larger than all its neighbours is associated with a local density minimum.}
 \item{Starting from a local minimum the algorithm appends neighbour cells with higher density (i.e. smaller size) than the minimum cell. The process stops when no more higher density cells are found. This procedures creates basins (called zones), that are depressions in the density field, which could be identified as sub-voids already.} 
\item{Finally, in order to make sense of the nesting of smaller voids into big voids, a watershed transform \citep[see e.g.][]{platen2007}  is applied to join the basins together and define a hierarchy of voids and sub-voids starting with the deepest basins.  Sub-voids are merged and nested into larger voids if the density of the ridge in between them is lower than the $20\%$ of the density of the Universe.  A number is assigned to each void with 0 meaning it is the deepest (i. e. the parent void) and (e.g.) 1 or 2 referring to sub-voids at different levels in the hierarchy.The TreeLevel parameter can be used to filter on the hierarchical properties of the voids  \citep[see e.g.][]{Lavaux2012}.}
\end{itemize}

After the so-described void-finding procedure is concluded, each void is assigned an effective radius $r_{v}^{\rm eff}$ that is equal to the radius of a sphere with a volume identical to the total void volume. Then centres of 3D  {\VIDE} voids are defined as volume-weighted barycenters of all the Voronoi cells that make up the given void.

We note that the possible elongation properties of {\ZOBOV}/{\VIDE} voids identified in photo-$z$ samples have also been investigated by \cite{granett2015} using overlapping tracer with accurate spectroscopic redshift information as ground truth. Then \cite{Fang2019} reconstructed the average shape of the DES Y1 and MICE {\VIDE} voids we also use in this study and reported a significant line-of-sight elongation (with an axis ratio of about 4) due to photo-$z$ errors. They concluded, however, that individual voids are not necessarily more elongated but a selection bias in orientation aligned with our line-of-sights breaks the isotropy. Relatedly, \cite{Cautun2018} argued that tunnel-like structures provide better signal-to-noise compared to spherical voids of the same angular size, and therefore this property of our {\VIDE} voids is not a disadvantage.

\subsection{Cosmic void properties in the MICE galaxy mocks}\label{sec:void_cat}

We note that the definition of effective radius of 3D {\VIDE} voids ($r_{v}^{\rm eff}$) is different than the radius definition of 2D voids ($r_v$) as we describe above. In particular, the void radius of {\VIDE} structures is defined as a turning point in the density profile's compensation around the voids, while the 2D void radius is simply a distance where the profiles reach the mean density. Similarly, the underdensity parameters are defined differently in the two void finders. Nevertheless the catalogues are internally consistent and their CMB lensing signals can meaningfully be compared to each other. We apply specific pruning methods to make 2D and {\VIDE} void catalogues more comparable, especially in number counts, and we provide a detailed description of these cuts in Section 4.

\subsubsection{2D voids}
We examine how potential systematic effects modify the resulting void populations. We compare the void parameter distributions for different tracer densities and various initial Gaussian smoothing applied to the density fields. Edge/mask effects may lead to different mean void properties because at survey boundaries the full extent of underdense regions around minima may not be captured with good precision. 

We run our 2D void finder using two different {\redmagic} samples as tracers. The {\redmagic} high-luminosity sample applies a stronger cut in luminosity ($L>1.5L_*$) which offers higher precision in photometric redshift. On the other hand, the {\redmagic} high-density sample has a more relaxed luminosity cut ($L>0.5L_*$), resulting in an increased galaxy density. We then further probe systematic effects by running the void finder on these two rather different samples using two different initial Gaussian smoothing scales, namely $10$ Mpc/h and $20$ Mpc/h. 

We compare the void catalogues in terms of three characteristic parameters of voids: distribution in physical size ($r_v$), distribution of mean density ($\bar{\delta}$) and distribution in central void density ($\delta_{1/4}$). We observe the following properties:

\begin{itemize}
\item Comparing the different resulting catalogues, a higher number of voids is detected when the tracer density is lower ({\redmagic} high-luminosity sample). \cite{Sutter_bias} found a different behaviour for {\VIDE} voids in simulations. Shot noise appears to drive these effects. In particular, a higher number of pixels are identified as 2D void centre candidates when the tracer density is lower, and the mean density might be reached more frequently, splitting voids up.

\item A larger smoothing scale decreases the total number of voids for both tracer densities, as the role of shot noise is reduced.

\item The mean void radius is shifted towards larger values for larger smoothings, as smaller voids merge into larger encompassing voids.

\item Small smoothing scales result in a larger fraction of deep voids, given the same tracer density. This feature is also related to shot noise properties.

\end{itemize}

When testing mask effects, we found that the voids identified using {\redmagic} tracers in the MICE octant have different properties compared to void properties of DES Y1-like survey patches inside the octant. We therefore decided to use the same mask as in the DES Y1 cosmological analysis \citep{wthetapaper} as this guarantees faithful comparison to the observed data. We consider two rotated positions of the Y1 mask with some overlap that is unavoidable inside the octant. Therefore, as a consistency test, we will study two MICE Y1-like void catalogues (MICE 1 and MICE 2; see Table \ref{table:voidnum} for more details).

\subsubsection{{\VIDE} voids}
Aiming at a different catalogue of voids from the same data set, we also run the {\VIDE} void finder on the MICE {\redmagic} photo-$z$ catalogue in the full octant, focusing on the high density sample of galaxies.

We find a total of 36115 voids using this 3-dimensional algorithm. The {\VIDE} algorithm provides various output parameters to characterise the voids. We judge that the most important parameters for our CMB lensing study are the effective radius ($r_{v}^{\rm eff}$), density contrast ($r$), and the ${\rm TreeLevel}$ \citep[for details see e.g.][]{zobov,sutter2015}.

Unlike for 2D voids, we find no significant difference in {\VIDE} void properties (such as radius, central underdensity, and redshift distribution) when using Y1-like mask patches or a full octant mask in MICE. This agrees with the findings of \cite{Pollina2018}. We therefore consider all voids in the MICE octant for our stacking tests, i.e. a factor of $\sim5$ more voids than in a Y1 patch (see also Table \ref{table:voidnum} for void number count comparisons).

In our empirical tests, we found that a $r_{v}^{\rm eff} > 35~{\rm Mpc/h}$ limit in radius effectively removes small voids that tend to live in overdense environments.The positive central $\kappa$ imprint of these small voids decreases the negative stacked $\kappa$ signal inside the void radius, bringing the signal closer to zero thus harder to detect. We also found that an additional cut that removes the least significant voids below the $1\sigma$ extremeness level ($r > 1.22$) \citep{zobov} is helpful to eliminate voids with less negative central imprints and remaining larger voids with positive central imprints. While these choices are subject to further optimisation, we use them in the present analysis in order to test a different definition using a robust and clean {\VIDE} sub-sample.

Finally, we apply a cut with ${\rm TreeLevel=0}$ to only keep voids which are highest in the hierarchy, i.e. do not overlap with sub-voids. These three conditions result in a set of voids that is a very conservative subset of the full catalogue. However, such a pruned catalogue with clean expected CMB $\kappa$ imprints is sufficient for providing an alternative for our main analysis with 2D voids.

\begin{table}
\centering

{\bf High luminosity (HL)}
\\
\begin{tabular}{@{}cccc}
\hline
\hline
Smoothing & DES Y1 & MICE 1 & MICE 2 \\
\hline
10 Mpc/h & 1218 & 1158 & 1219 \\
20 Mpc/h & 411 & 364 & 400
\\
\hline
\hline
\end{tabular}
\\

{\bf High density (HD)}
\\
\begin{tabular}{@{}cccc}
\hline
\hline
Smoothing & DES Y1 & MICE 1 & MICE 2 \\
\hline

10 Mpc/h & 427 & 421 & 420 \\
20 Mpc/h & 122 & 85 & 106
\\
\hline
\hline

{\VIDE} & DES Y1 & MICE &  \\
\hline
All & 7383 & 36115 & \\
Pruned & 239 & 1687 & \\
\hline
\hline
\end{tabular}

\caption{We list the numbers of 2D voids identified in two Y1-like MICE patches vs. in DES Y1 data. We also provide void number counts for {\VIDE} voids for the full MICE octant and for the DES Y1 data set, with and without pruning cuts that we consider in our measurements.}\label{table:voidnum}
\end{table}

\subsection{DES Y1 catalogues compared to simulations}
In the light of the simulated stacking measurements using the MICE $\kappa$ map, we aim to measure the DES Y1 voids $\times$ {\it Planck} CMB $\kappa$ signal. We thus use the observed {\redmagic} catalogues from DES Y1, presented in \ref{sec:void_cat}, to construct void catalogues with the different tracer densities and initial smoothing scales.

Figure \ref{fig:SIM_DATA_void_compa} shows a comparison of the observed and simulated 2D void catalogues. We report a very good agreement in terms of sizes, central density, and mean density for both MICE Y1-like patches when they are compared to DES Y1 data. We find that the simple two-sample Kolmogorov-Smirnov (KS) histogram consistency tests \citep{Kolmogorov,Smirnov:1948:TEG} suggest that, in general, high luminosity samples are in slightly better agreement (see Table \ref{table:voidnum}). However, the overall agreement is sufficient (with KS test p-values ranging from 0.28 to 0.97), thus we aim to test the consistency of simulations and observations for all void catalogue versions.

We also find good agreement between void properties of the simulated and observed catalogues using the {\VIDE} algorithm on the DES Y1 \redmagic\ high density sample. We identify a total of 239 voids in DES Y1 data considering the selection cuts explained above. This is a very conservative cut on the total of 7383 voids in the DES Y1 {\VIDE} catalogue that also includes smaller and less significant voids. Our primary goal with this work was to offer a robust alternative to 2D voids, and we thus leave the further optimisation of the {\VIDE} sample for future work.

\begin{figure*}
\begin{center}
\includegraphics[width=185mm]{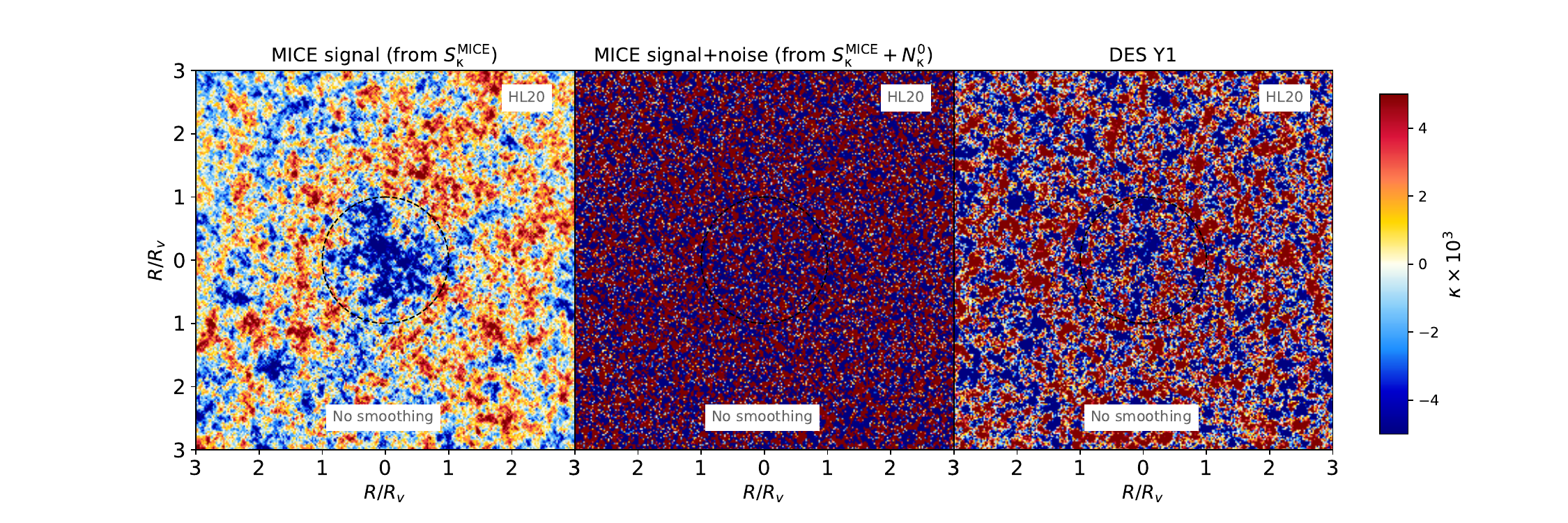}
\includegraphics[width=185mm]{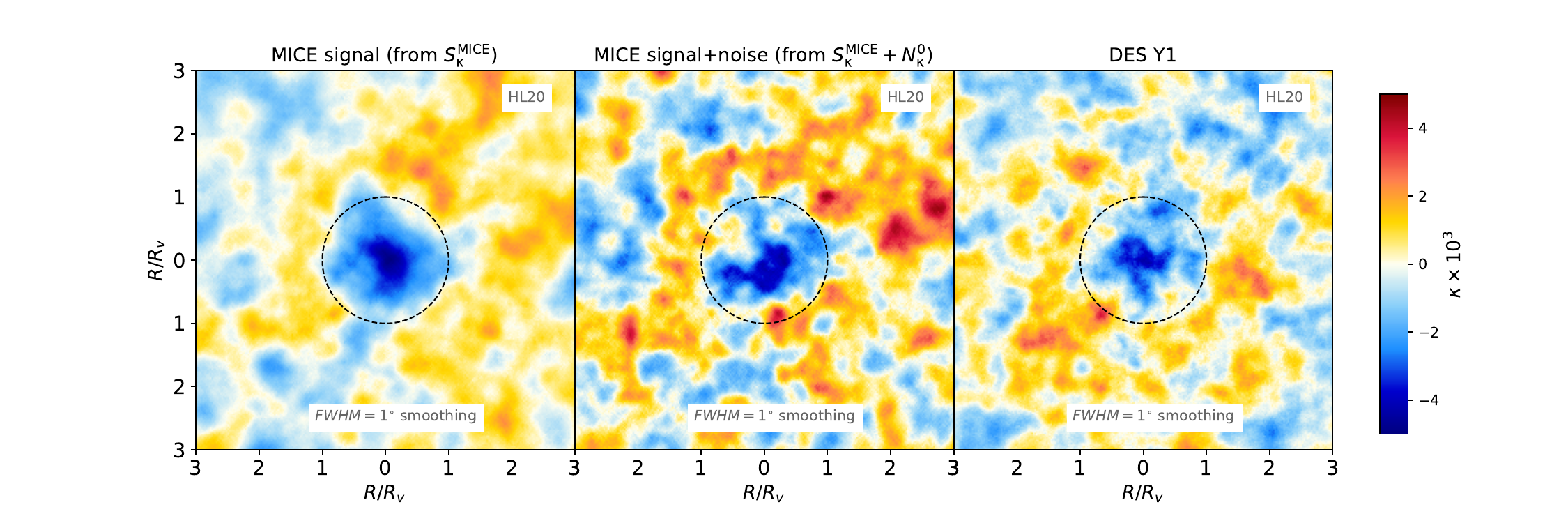}
\includegraphics[width=185mm]{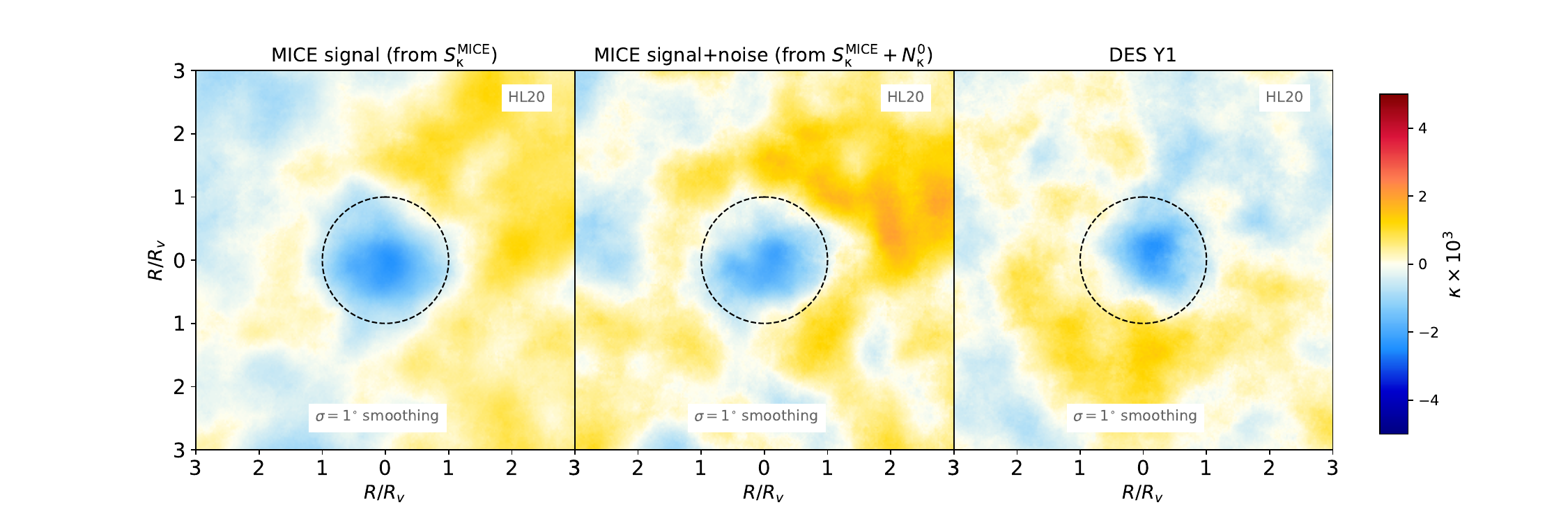}
\end{center}
\caption{Simulated signal-only stacked $\kappa$ images from MICE (left) in comparison to noise-added versions (centre) and observed DES Y1 stacked results (right) for the HL20 version of 2D voids. All versions of our results are displayed, without smoothing (top) and with \textit{FWHM}$=1^{\circ}$ (middle) or $\sigma=1^{\circ}$ (bottom) Gaussian smoothings are used. The re-scaled void radius $R/R_{v}=1$ is marked by the dashed circles. We identify important trends with changing smoothing scales but overall report good consistency between data and simulations.}
\label{fig:stacked_2D}
\end{figure*}

\begin{figure*}
\begin{center}
\includegraphics[width=185mm]{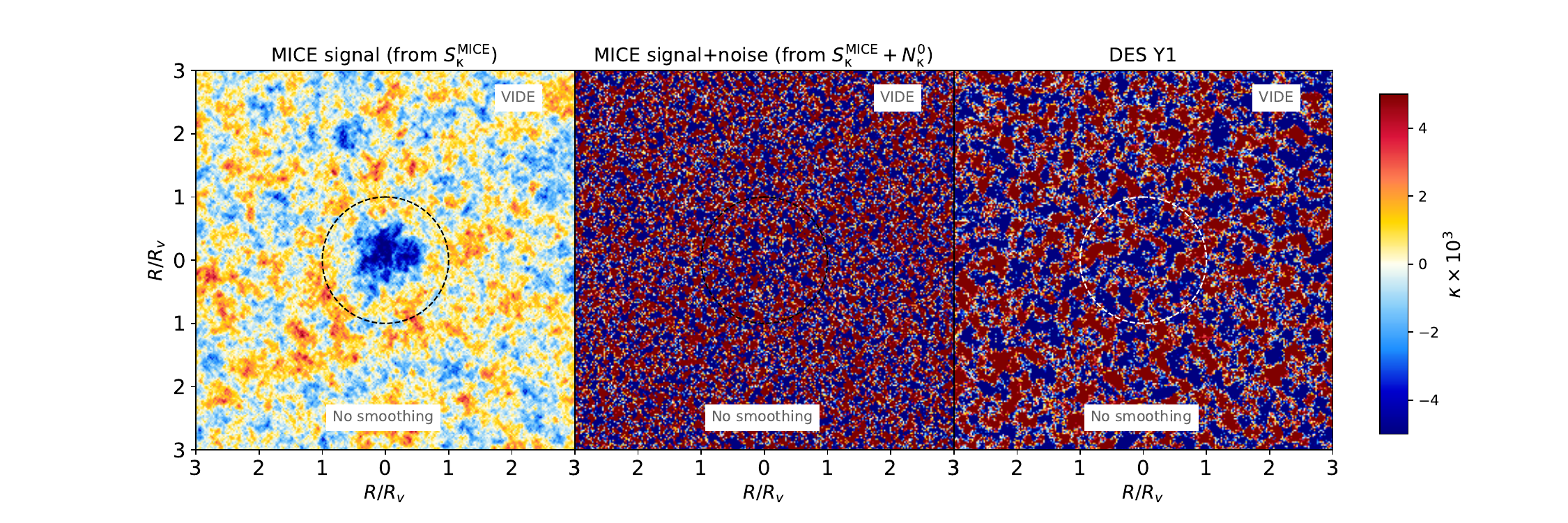}
\includegraphics[width=185mm]{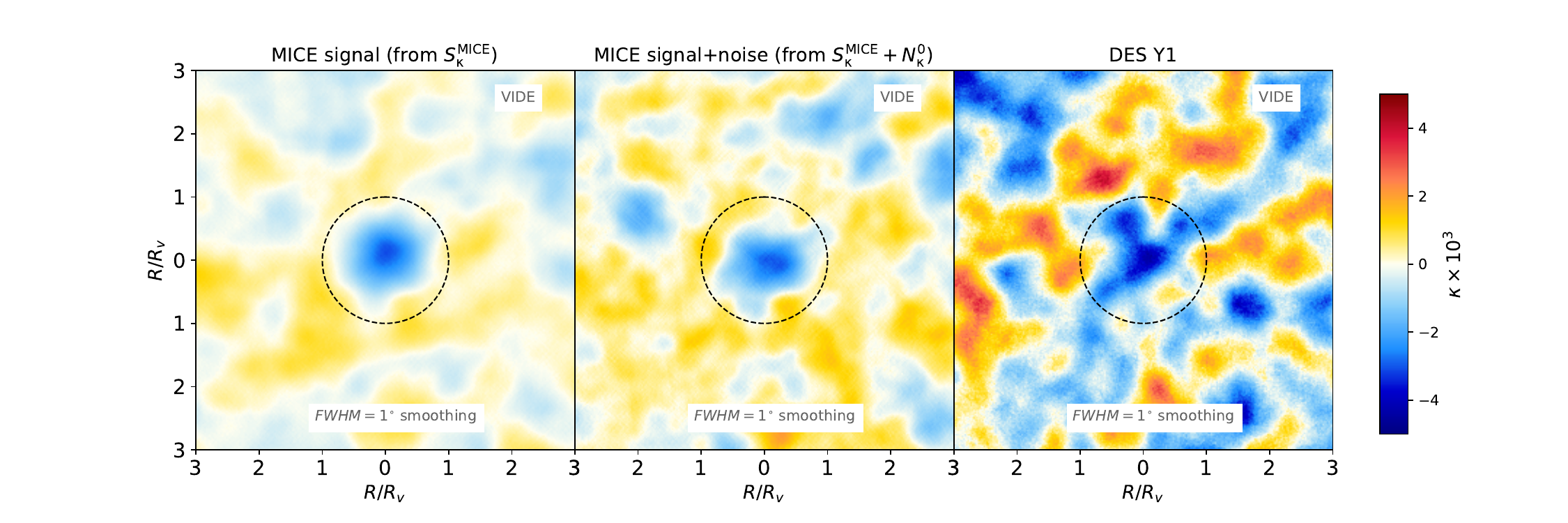}
\includegraphics[width=185mm]{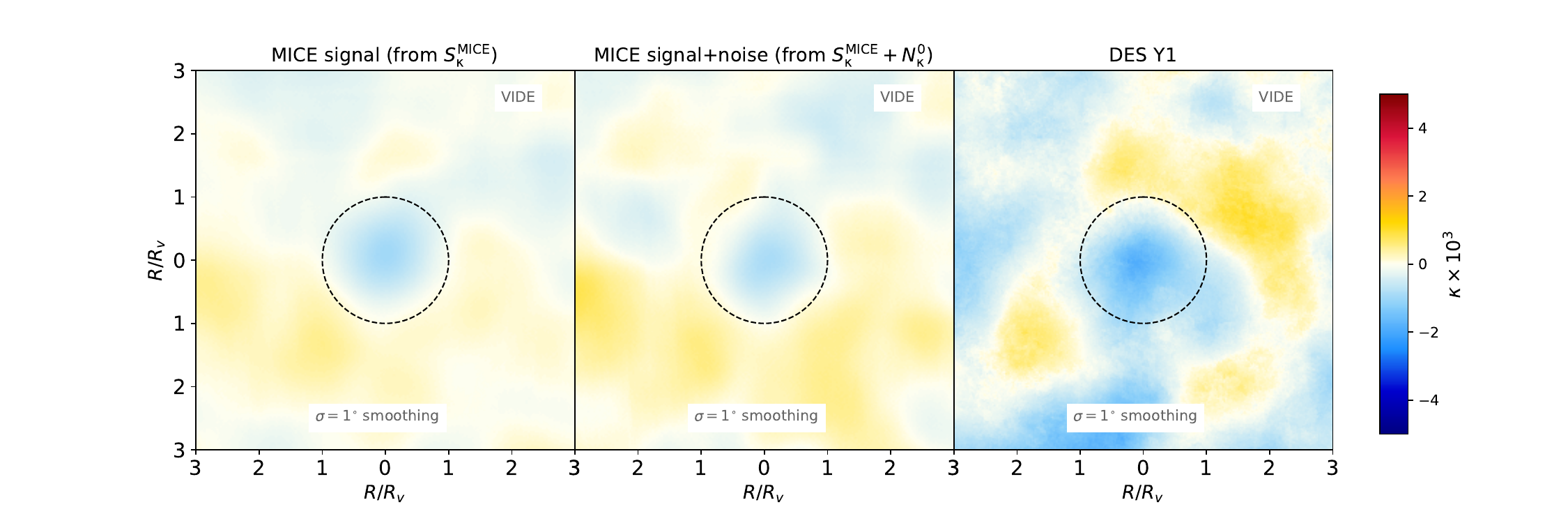}
\end{center}
\caption{Same as Figure \ref{fig:stacked_2D} except we replace the 2D void sample with {\VIDE} voids.}
\label{fig:stacked_ZOBOV}
\end{figure*}

\section{Simulated cross-correlation analyses}
\label{Section4}
\subsection{Stacking $\kappa$ maps on void positions}

The CMB lensing imprint of single voids is impossible to detect \citep[see e.g.][]{Krause2013}. We therefore apply an averaging method using cutouts of the CMB map at void positions \citep[see e.g.][and reference therein]{Kovacs2016}.
This stacking procedure can be described with the following steps:
\begin{itemize}
    \item we define a catalogue of voids. We also select subgroups in radius and density bins to probe their specific imprint type.
    \item we re-scale the angular size of voids to measure distances in dimensionless $R/R_{v}$ units. We use a patch size enclosing $5$ re-scaled void radii to possibly detect the lensing imprint of void surroundings, such as matter overdensities around voids, i.e. compensation walls \citep{Hamaus2014}.
    \item we probe the effect of a Gaussian smoothing on the noise properties of the stacked images using different filter sizes applied to the CMB convergence map (not in the re-scaled images).
    \item we stack using three different strategies: without smoothing; using a full width at half maximum value \textit{FWHM}$=1^{\circ}$; and with a standard deviation $\sigma=1^{\circ}$ (equivalent to \textit{FWHM}$=2.355^{\circ}$) to reduce the noise of the measurement. A more optimized analysis could use filters matching the shape of the expected signal to maximize $S/N$ \cite[see][for a similar analysis]{NadathurCrittenden2016}. \item we found that \textit{FWHM}$=1^{\circ}$ is a good compromise as it efficiently removes fluctuations from very small scales (compared to the typical void size) but it practically preserves the signal itself (see Figure \ref{fig:stacked_ZOBOV} for details).
    \item we cut out the re-scaled patches of the CMB convergence map centred at the void centre position using \texttt{healpix} tools \citep{Healpix}. This allows us to have the same number of pixels by varying the resolution of the images according to the particular void angular size.
    \item we then stack all patches and measure the average signal in different concentric radius bins around the void centre.
\end{itemize}

As we use full-sky MICE $\kappa$ maps but only consider smaller DES Y1-like patches, we also measure the mean $\kappa$ values in the masked area and remove this bias from the profiles to account for possible large-scale fluctuations that a DES Y1-like survey is affected by. From the {\it Planck} data, we also remove the mean $\kappa$ value measured in the DES Y1 footprint.
We do not apply any other filtering in the stacking procedure such as exclusion of large-scale modes up to $\ell<10$ \citep[see][for related results]{cai}.

\subsection{Simulated analyses with noise in the $\kappa$ map}
\label{errors}

An important source of the measurement uncertainties are the random instrumental noise in the {\it Planck} data. In order to model observational conditions, we generate 1000 {\it Planck}-like noise map realisations using the noise power spectra released by the {\it Planck} team \citep{Planck2018_lensing}. We first check how the detectable signal fluctuates around the true signal without rotating the MICE lensing map ($S_{\rm \kappa}^{\rm MICE}$) in alignment with void positions. In this test we add simulated noise contribution maps ($N_{\rm \kappa}^{i}$) to the same non-rotated MICE $\kappa$ (signal-only) map in 1000 random realisations. We find significant fluctuations in the signal in the presence of {\it Planck}-like noise but no evidence for biases when considering noisy data. Figure \ref{fig:stacked_2D} shows how the signal-only ($S_{\rm \kappa}^{\rm MICE}$) and noise-added ($S_{\rm \kappa}^{\rm MICE} + N_{\rm \kappa}^{0}$) MICE images compare for a given noise realisation in the case of 2D voids.

We note, however, that the total error of the stacking measurement also has a contribution from random fluctuations in the stacked signal map itself. This sub-dominant contribution is about half the magnitude of the instrumental $\kappa$ noise based on comparisons of fluctuations in random stacking measurements using the signal-only MICE map $S_{\rm \kappa}^{\rm MICE}$ or $N_{\rm \kappa}^{i}$ noise maps. This second error is, at least in part, due to the complicated overlap structure of voids themselves along the line-of-sight, overlap with their neighbour voids in the same redshift slice, and also the limited number of available voids in a DES Y1 observational setup. These result in imperfect imprints compared to a hypothetical mean signal of several isolated voids.

Then, to account for both the above sources of error in the void-$\kappa$ cross-correlation measurement, we first create 1000 noise-added $S_{\rm \kappa}^{\rm MICE} + N_{\rm \kappa}^{i}$ maps. In this case, we randomly rotate $S_{\rm \kappa}^{\rm MICE}$ and estimate the measurement errors with 1000 runs (void positions and the $S_{\rm \kappa}^{\rm MICE}$ map are not aligned). However, as the rotated MICE maps may overlap in our 1000 random rotations affecting the estimation of the covariance, we consider an alternative strategy to estimate the measurement errors. We measure the power spectrum of the noiseless full-sky MICE $\kappa$ map $S_{\rm \kappa}^{\rm MICE}$ using the \texttt{anafast} routine of \texttt{healpix}. Then, given the same power spectrum, we create 1000 random map realisations using \texttt{synfast}. We then add our $N_{\rm \kappa}^{i}$ noise map realisations to these different $S_{\rm \kappa}^{i}$ MICE-like lensing map realisations, and thus eliminate the possible correlations between random realisations due to rotation of the MICE map. Finally, we stack the 1000 noisy random maps on void positions, and, as in the MICE and DES Y1 measurements of the imprint signals, we also remove the mean $\kappa$ map value inside the DES Y1 survey mask area. Our additional tests show that the removal of this monopole $\kappa$ bias term reduces the overall errors on the $A$ lensing amplitude by about $10\%$.

We note that while simulated and observed void catalogues are in good agreement (see Figure \ref{fig:SIM_DATA_void_compa}), we use the observed DES Y1 void catalogues for the estimation of the errors to ensure that the overlap structure or any other correlation between voids is fully realistic. We find that the above error estimation methods give consistent results, but the second \texttt{synfast}-based method provides a few per cent larger error bars. This is intuitively expected, since slightly more randomness is added to the stacking process by using independent $\kappa$ maps instead of rotation of a single one. We therefore consider these more conservative \texttt{synfast}-based errors in our covariance estimation process.

For all void catalogues, we repeat all measurements for our three different $\kappa$ smoothing strategies: no smoothing, and two Gaussian smoothings with \textit{FWHM}$=1^{\circ}$ and $\sigma=1^{\circ}$. Figure \ref{fig:stacked_2D} demonstrates how different smoothings of the $\kappa$ maps affect the results. In Figure \ref{fig:stacked_2D} we also preview the results from stacking measurements using a DES Y1 2D void catalogue to show the reasonable agreement between noise-added simulations and observed data. Other versions of the void catalogue showed consistent results. Figure \ref{fig:stacked_ZOBOV} presents our findings on alternative {\VIDE} void catalogues in MICE and in DES Y1. We find imprints comparable to 2D void results for our very conservative subset.

\subsection{Amplitude fitting}

In our DES Y1 analysis we wish to perform a template fitting algorithm using the simulated radial $\kappa$ profiles extracted from MICE stacking analyses. As a measure of the signal-to-noise (S/N) of simulated and observed signals given the  measurement errors and their covariance, we aim to constrain an amplitude $A$ (and its error $\sigma_{\rm A}$) as a ratio of DES Y1 and MICE signals using the full profile up to $R/R_{v}=5$ in 16 radial bins. We expect $A=1$ if the DES Y1 and MICE $\Lambda$CDM results are in close agreement and we aim to test this hypothesis. In the DES Y1 analysis, we fix the shape of the stacked convergence profile to that calibrated from the MICE simulation. See e.g. \cite{Kovacs2019} for a similar analysis with DES voids.

As detailed above, we estimate the covariance using 1000 different {\it Planck}-like noise simulations (that dominate the measurement errors), and we also add a randomly generated CMB lensing map with MICE-like power spectrum to estimate the full error. We then invert the covariance matrix and correct our estimates by multiplying with the Anderson-Hartlap factor $\alpha=(N_{\rm randoms}-N_{\rm bins}-2)/(N_{\rm randoms}-1)$ \citep{hartlap2007}. Given our measurement configuration, this serves as a small ($\approx2\%$) correction.

\begin{figure*}
\begin{center}
\includegraphics[width=180mm]{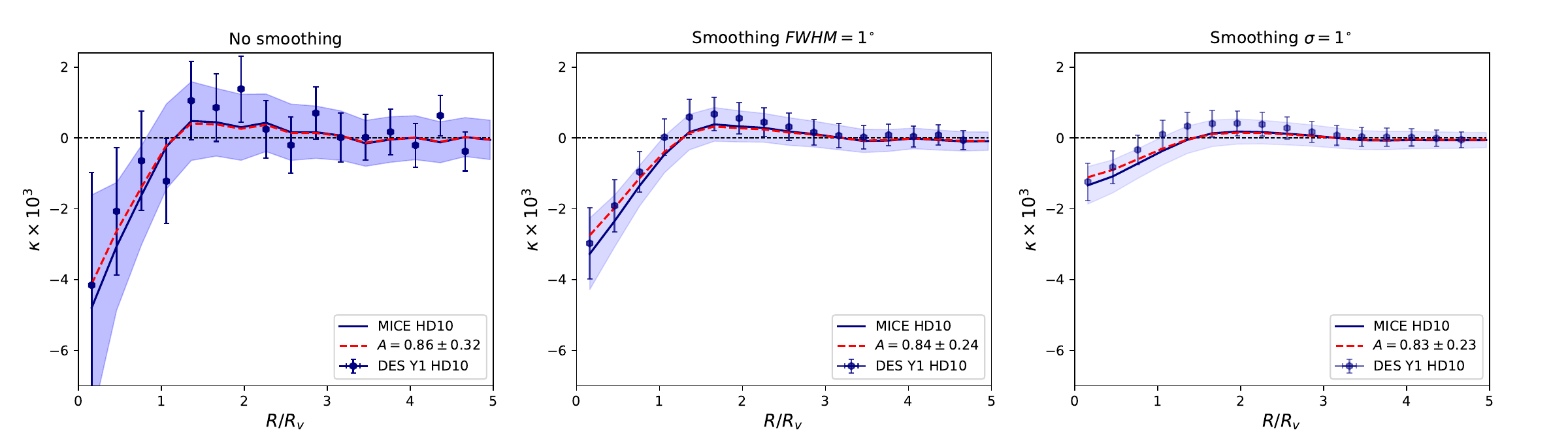}
\includegraphics[width=180mm]{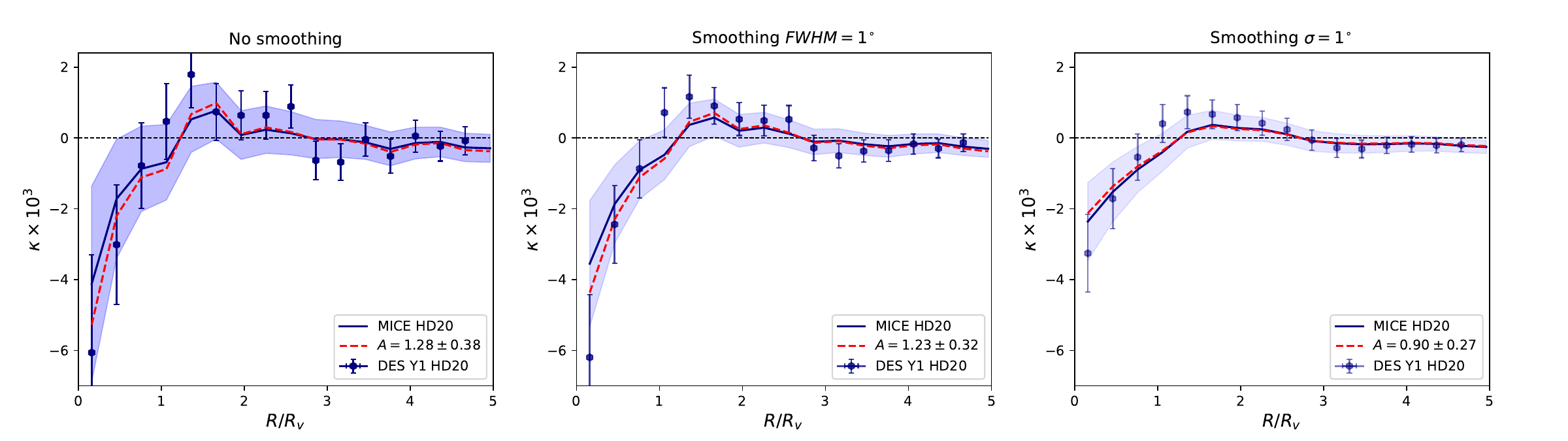}
\includegraphics[width=180mm]{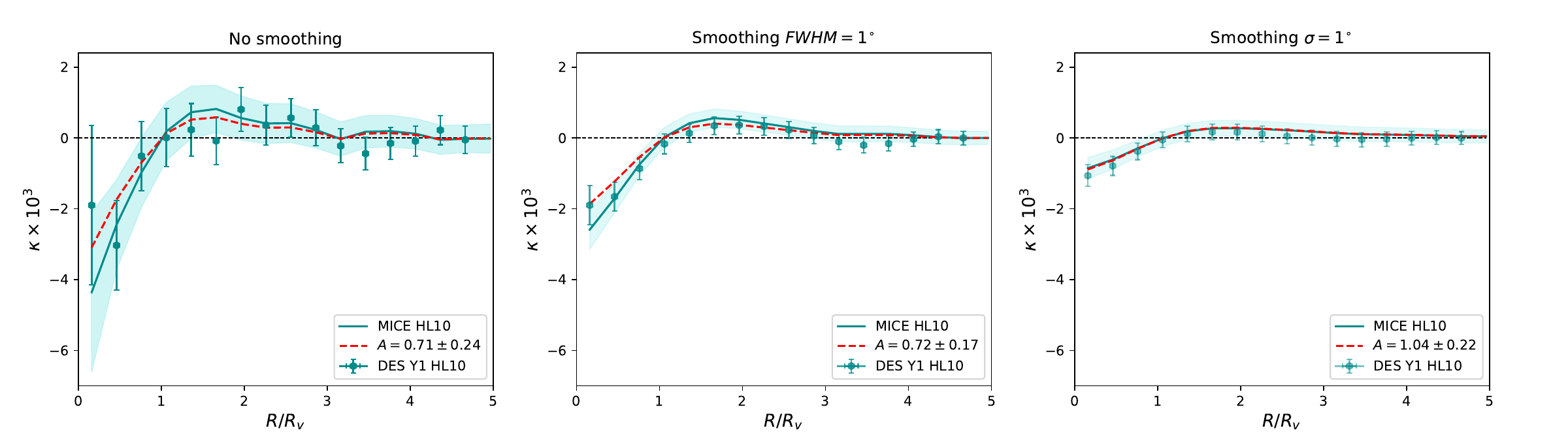}
\includegraphics[width=180mm]{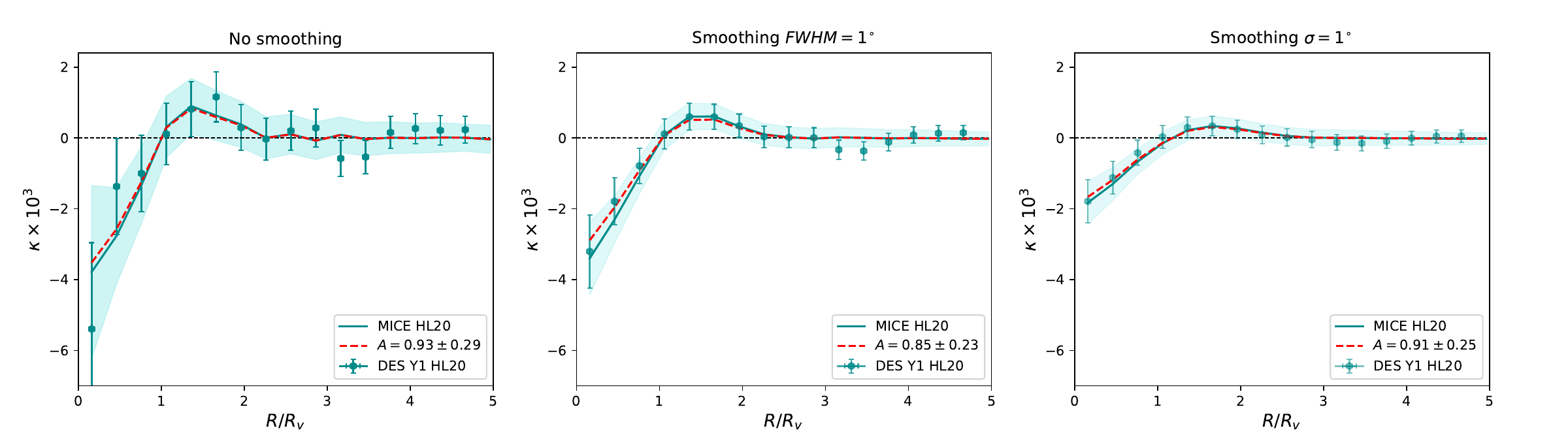}
\end{center}
\caption{Comparison of the radial $\kappa$ imprint profiles of 2D voids in the MICE simulation and in DES Y1 data. We show results based on all three $\kappa$ map smoothing strategies, including no smoothing (left), \textit{FWHM}$=1^{\circ}$ smoothing (middle), and $\sigma=1^{\circ}$ smoothing (right). For completeness, we present the imprints for all 2D void catalogue versions including HD10, HD20, HL10, and HL20 from top to bottom. Dashed red profiles mark the best fitting MICE templates considering the DES measurements.}
\label{fig:stack_2Dvoids_prof}
\end{figure*}

To constrain the $A$ amplitude, we then evaluate a statistic
\begin{equation}
{\chi}^2 = \sum_{ij} (\kappa_{i}^{\rmn{DES}}-A\kappa_{i}^{\rmn{MICE}}) C_{ij}^{-1} (\kappa_{j}^{\rmn{DES}}-A\kappa_{j}^{\rmn{MICE}})
\end{equation}
where $\kappa_{i}$ is the mean lensing signal in the radius bin $i$, and $C$ is the covariance matrix defined above. We perform such a ${\chi}^2$ minimization for all void catalogue versions and smoothing strategies using the corresponding data vectors and covariances.

\subsection{Optimization of the measurement}\label{sec:optimization}

The imprint of voids on the CMB lensing maps depends on their properties. \cite{nadathur2017} showed that simulated cosmic voids, identified with the {\ZOBOV} methodology (similar to {\VIDE}), trace the peaks of the underlying gravitational potential differently given different density, size, and environment \citep[see also][]{cai}]. They reported that voids can be grouped based on a combined density-radius observable to have distinct lensing profiles. In particular, they found that the combination of all sub-populations gives an average profile that is closer to zero at all scales, i.e. harder to detect. For instance, stacked $\kappa$ images of voids-in-voids are less negative in their centre, while voids-in-clouds show a more pronounced compensation. The overall significance of the measurement can therefore be improved if the distinct imprints of different void types are measured separately and a combined significance analysis is performed. These findings appear to be robust against changing the galaxy tracer sample but have not yet been tested in photo-$z$ void studies. We thus cannot blindly follow these pruning strategies in our methodology.

\subsubsection{2D voids}

While 2D voids are different in their nature than 3D voids, we aim to explore the possible optimisation of the void catalogue by pruning in a similar manner. We therefore perform the stacking measurement for subsets of our 2D void catalogues for both tracer densities and two different initial density smoothing scales. 

The S/N is first measured in stacked images using individual bins in void radius and underdensity, indicating how sub-classes of voids contribute to the total detection significance. Similarly, we also stack cumulatively, i.e. gradually making use of all the voids in the sample by adding more and more voids from bins of $r_{\rm v}$ and  $\bar{\delta}$, indicating which portion of the radius-ordered and density-ordered data provides the highest detection significance. We make the following observations based on these optimisation efforts:
\begin{itemize}
    \item medium size voids of radii $40$ ${\rm Mpc/h}< r_{v} <80$ ${\rm Mpc/h}$ account for most of the observable lensing signal. The magnitude of their lensing imprint is the highest and they are the most numerous subgroup in the void catalogue that results in smaller uncertainties.
    \item splitting the void catalogue based on the mean underdensity in voids, we find that voids with $-0.2<\bar{\delta}<-0.1$ carry most of the observable signal. These are rather shallow void structures but they are the most numerous that naturally result in higher statistical precision in the stacking measurements.
    \item while approximately two thirds of the S/N is contained inside the void radius ($R/R_{v}<1$) and in the close surroundings ($1<R/R_{v}<2$), measuring the cumulative S/N up to ($R/R_{v}=5$) does increase the detectability and provides a way to test convergence to zero signal at large radii.
    \item the highest S/N is achieved by stacking all voids, even if some voids are expected to contribute with less pronounced signal and higher noise at small scales \citep[see][for a counter-example in the case of ISW imprints]{Kovacs2016}.
   \end{itemize}
   
\begin{figure*}
\begin{center}
\includegraphics[width=180mm]{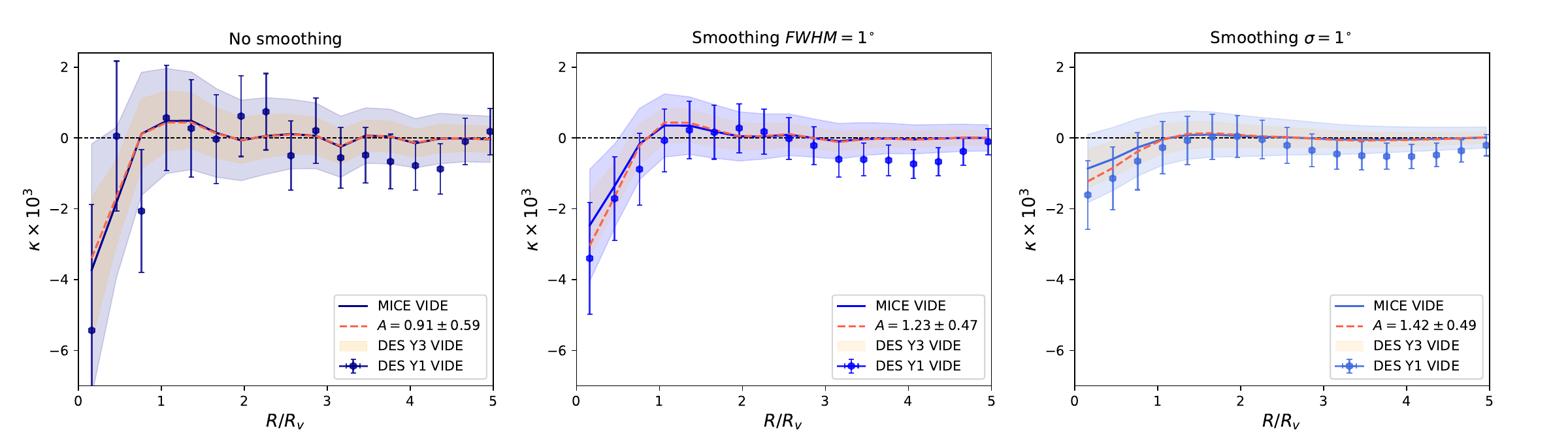}
\end{center}
\caption{We compare the radial $\kappa$ imprint profiles of {\VIDE} voids in the MICE simulation and in DES Y1 data. We show results based on all three $\kappa$ map smoothing strategies. Dashed red profiles mark the best fitting MICE templates to the DES measurements. We also mark the expected errors for the Year-3 DES data set that we wish to use in the future to extend this analysis (orange shaded areas around the MICE signals).}
\label{fig:stack_ZOBOV_prof}
\end{figure*}

In terms of different tracer density and smoothing, the highest S/N is found when using the high luminosity catalogue with $10$ Mpc/h smoothing (HL10). We note that such a result is not unexpected, given the wider redshift range and the larger fraction of deep voids in the case of the HL sample (see Figure \ref{fig:SIM_DATA_void_compa}).

We estimate $S/N=4.2$ for the case of no $\kappa$ map smoothing, while we find an even higher $S/N=5.9$ and $S/N=4.6$ for Gaussian smoothings using \textit{FWHM}$=1^{\circ}$ and $\sigma=1^{\circ}$, respectively. We use $S/N$ and $A/\sigma_{\rm A}$ interchangeably to refer to the signal-to-noise throughout the paper. We consider a DES Y1 measurement configuration and resulting errors and a MICE $\Lambda$CDM signal ($A=1$) of the simulated 2D voids.

Nevertheless, all measurement configurations show moderately significant $S/N \gtrsim 3$ CMB lensing signals for voids in a survey such as DES Y1, and thus we will measure the corresponding observed lensing imprint of all DES void catalogues and smoothing versions. See again Figure \ref{fig:stacked_2D} for details.

We note that the main results above are based on the full void sample with a variety of redshifts in $0.2<z<0.7$. For completeness, we also performed a simple redshift binning test for voids of size $20$ $ {\rm Mpc/h} < r_{v} < 70$ ${\rm Mpc/h}$. We found no clear evidence for redshift evolution in their CMB lensing profile.

\subsubsection{{\VIDE} voids}

Because in this paper we consider {\VIDE} voids as a consistency test, we do not formally optimise the signal-to-noise for the {\VIDE} void sample. Relatedly, we do not have a single recipe for pruning parameters in the presence of photo-$z$ errors for 3D voids. Nevertheless, as explained in Section 3.3.2, we apply various pruning cuts in order to ensure a detectable CMB lensing signal in the MICE simulation and therefore also in DES Y1 data (see Figure \ref{fig:stacked_ZOBOV}). These cuts result in 1687 {\VIDE} voids in the MICE octant to be used in the stacking measurement, and 239 voids in the DES Y1 \redmagic\ high density data. We present a comparison with 2D void types in Table \ref{table:voidnum}, finding good consistency in void number counts.

Overall, we find $S/N=1.7$ for the case of no $\kappa$ map smoothing, while $S/N=2.1$ and $S/N=2.0$ for Gaussian smoothings using \textit{FWHM}$=1^{\circ}$ and $\sigma=1^{\circ}$, respectively. In these tests, we again consider a MICE $\varLambda$\textit{CDM} imprint signal ($A=1$) and a DES Y1 measurement configuration and resulting errors ($\sigma_{\rm A}$) of the simulated {\VIDE} voids.

We note that our pruning cuts in fact remove most of the voids from the original catalogue; thus the {\VIDE} catalogue may promise higher $S/N$ with further optimisation. However, for our purposes of studying a sample complementary to the 2D void analysis the sample defined above is adequate. We leave the optimisation of {\VIDE} catalogues for CMB lensing measurements for future work, including tests of {\VIDE} voids in high luminosity tracer catalogues that appear more promising for the 2D void definition.

\section{Results for observations: DES Y1 $\times$ \emph{Planck}}
\label{Section5}

We measure the stacked imprint of DES Y1 voids with the same methodology and parameters as in the case of the MICE mock. Together with the MICE results, the stacked $\kappa$ images of the DES Y1 void catalogues are shown in Figures \ref{fig:stacked_2D} and \ref{fig:stacked_ZOBOV} for 2D and {\VIDE} voids, respectively. We find good consistency between simulations and observations for all void definitions, smoothing strategy, and tracer density.

We then use the stacked images to calculate a radial $\kappa$ imprint profile in order to quantify the results, relying on the noise analysis we introduced above. We present these results below and provide a detailed description of our constraints on the $A$ amplitude of DES Y1 and MICE void lensing profiles.

\begin{table}
\centering
{\bf No smoothing}
\\
\begin{tabular}{@{}cccccc}
\hline
\hline
Catalogue & {\VIDE} & HD10 & HD20 & HL10 & HL20 \\
\hline
MICE & 1.69 & 3.12 & 2.63 & 4.16 & 3.45 \\
\hline
DES Y1 & 1.54 & 2.68 & 3.40 & 2.94 & 3.15 \\
\hline
\end{tabular}

{\bf \textit{FWHM}$=1^{\circ}$ smoothing}
\\
\begin{tabular}{@{}cccccc}
\hline
\hline
Catalogue & {\VIDE} & HD10 & HD20 & HL10 & HL20 \\
\hline
MICE & 2.12 & 4.16 & 3.12 & 5.88 & 4.35 \\
\hline
DES Y1 & 2.61 & 3.46 & 3.80 & 4.13 & 3.70 \\
\hline
\end{tabular}

{\bf $\sigma=1^{\circ}$ smoothing}
\\
\begin{tabular}{@{}cccccc}
\hline
\hline
Catalogue & {\VIDE} & HD10 & HD20 & HL10 & HL20 \\
\hline
MICE & 2.04 & 4.34 & 3.70 & 4.55 & 4.00 \\
\hline
DES Y1 & 2.89 & 3.55 & 3.38 & 4.74 & 3.62 \\
\hline
\end{tabular}
\caption{\label{table2} Signal-to-noise ratios ($A / \sigma_{\rm A}$) are listed for all measurement configurations using MICE and DES Y1 signals. We compare three different smoothing strategies and five void catalogue versions.}
\end{table}

\begin{figure*}
\begin{center}
\includegraphics[width=180mm]{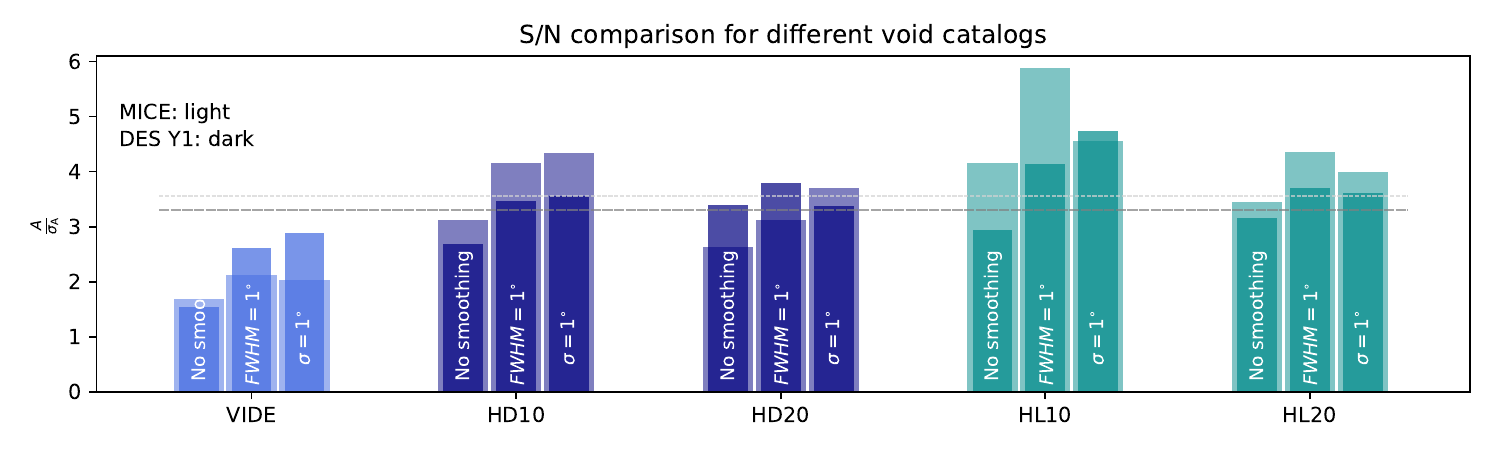}
\end{center}
\caption{We provide a detailed comparison of measurement significance in the form of $A / \sigma_{\rm A}$. The conservative {\VIDE} sample also provides useful consistency tests in agreement with our 2D analyses. The dashed horizontal lines mark the mean of the DES Y1 (dark) and the MICE (light) significances with values 3.31 and 3.55, respectively.}
\label{fig:summary}
\end{figure*}

\subsection{2D voids}

We continue our data analysis with the DES Y1 2D void catalogues that promised higher $S/N$ in our MICE analysis, where, recall, we forecasted $S/N\approx5$ for the high luminosity catalogue.

We compare the stacked images of the $\kappa$ imprints in the high luminosity catalogue with $20$ Mpc/h smoothing in the galaxy density map in Figure \ref{fig:stacked_2D} as a representative example of all 2D void results. A visual inspection shows good agreement between MICE and DES Y1 $\kappa$ imprints both in the centres and surroundings of the voids. We find consistency for all $\kappa$ smoothing strategies and report that similar conclusions can be drawn from stacked images from other void catalogue versions (see also Figure \ref{fig:stacked_ZOBOV}).

We then also measure the azimuthally averaged radial imprint profile in the stacked images to quantify the results. We present the results in Figure \ref{fig:stack_2Dvoids_prof} for all four 2D void catalogue versions HD10, HD20, HL10, and HL20. The shaded blue regions mark $1\sigma$ errors computed with 1000 random realisations of the stacking measurement on the MICE $\kappa$ map with {\it Planck}-like noise included, while the error bars around DES Y1 measurements show the same uncertainties for the DES data (by construction, we use the same covariance estimation methodology for MICE and DES data as explained in Section \ref{errors}). We observe a good general agreement in the sign and the shape of the observed and simulated profiles. Negative $\kappa$ values in the interior of voids plus an extended range of positive convergence in the surroundings. We note that the approximate convergence of the profiles to zero signal at large distance from the void centre is an important null test which proves that our method is not affected by significant additive biases.

We provide the $S/N$ ratios for all catalogue versions and analysis techniques in Table \ref{table2} and amplitudes with errors in Table \ref{table3}. We observe clear trends in the results, including a natural decrease of both errors and the signal itself if larger Gaussian $\kappa$ smoothing scales are applied to the CMB map. We see no evidence for significant excess signals or a lack of signal compared to simulations.

As demonstrated in detail in Figure \ref{fig:stack_2Dvoids_prof} for the case of 2D voids, the less promising DES Y1 void catalogue versions tend to robustly show signal-to-noise ratios of at least $S/N\approx3$. This is in good agreement with the mean of all MICE signal-to-noise estimates $S/N\approx3.5$. We compare these mean $S/N$ values to individual estimates in Figure \ref{fig:summary}. We find that the DES Y1 constraints on the $A$ amplitude typically favor values slightly lower than $A=1$, often with $A\approx0.8$, and this reduces the significance of our detections.

In particular, the highest signal-to-noise is expected for the HL10 sample with \textit{FWHM}$=1^{\circ}$ smoothing (based on the MICE analysis) with $S/N\approx5.88$. Using the DES Y1 catalogue we constrain $A\approx0.72\pm0.17$ and $S/N\approx4.13$, i.e. slightly lower than expected. In another promising configuration with the HL10 sample with $\sigma=1^{\circ}$ smoothing, we find $A\approx1.04\pm0.22$ and $S/N\approx4.74$, i.e. slightly higher than expected. Nevertheless we conclude that these results are consistent with expectations from MICE both in terms of amplitude and significance.

We note that our estimates of the stacked CMB $\kappa$ profile in the MICE mock are in good agreement with the simulated profile shapes and central amplitudes reported by \cite{cai} and \cite{nadathur2017} even though they used different void definitions and tracer catalogues.

\subsection{{\VIDE} voids}

In Figure \ref{fig:stack_ZOBOV_prof}, we present the profile measurement results for {\VIDE} voids for all three smoothing strategies. The profiles with error bars again indicate the signal-to-noise of the visually compelling imprints seen in the stacked images. We conclude that an \textit{FWHM}$=1^{\circ}$ smoothing offers the best chance to detect a signal. The detection reaches $S/N=2.6$ with $A\approx1.23\pm0.47$, given the DES Y1 survey setup, in good agreement with our predictions from the MICE mock (see more detailed comparisons of expected and measured $S/N$ in Figure \ref{fig:summary}). We find that the best-fit amplitudes are all consistent with the expectation $A=1$ from the MICE simulation.

As a forecast, in Figure \ref{fig:stack_ZOBOV_prof} we over-plot the expected error bars for the upcoming DES Y3 release that will offer a better chance to measure the void CMB lensing signal of DES voids even with a conservatively pruned {\VIDE} catalogue. We expect roughly two times smaller error bars given the approximately four times larger survey area. This translates to an expected $S/N\approx4.5$ detection for identically selected but more numerous DES Y3 {\VIDE} voids.

\begin{figure*}
\begin{center}
\includegraphics[width=175mm]{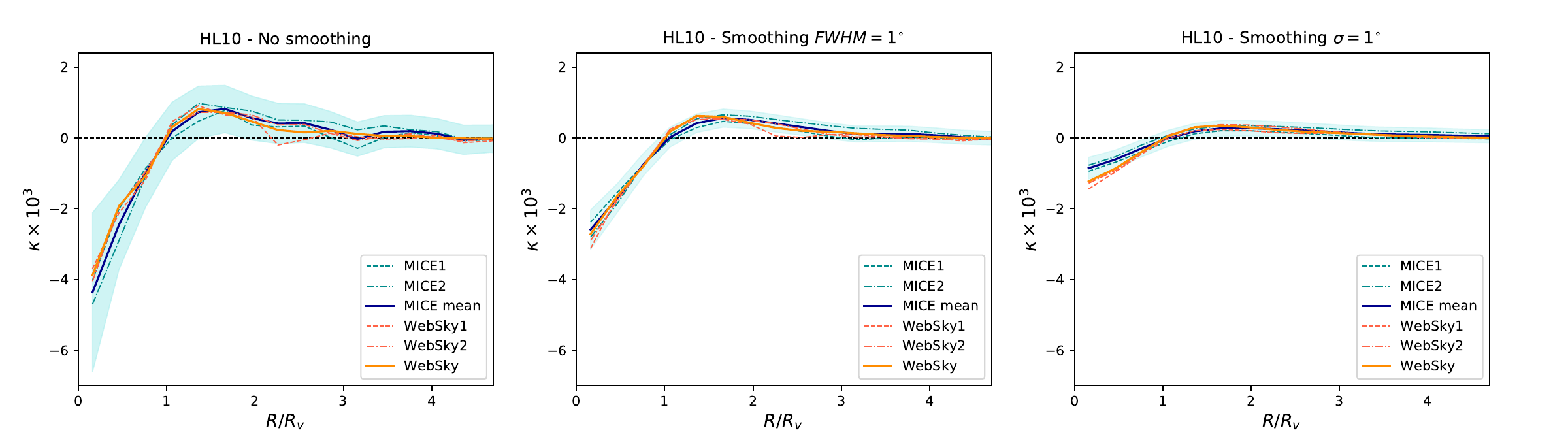}
\includegraphics[width=175mm]{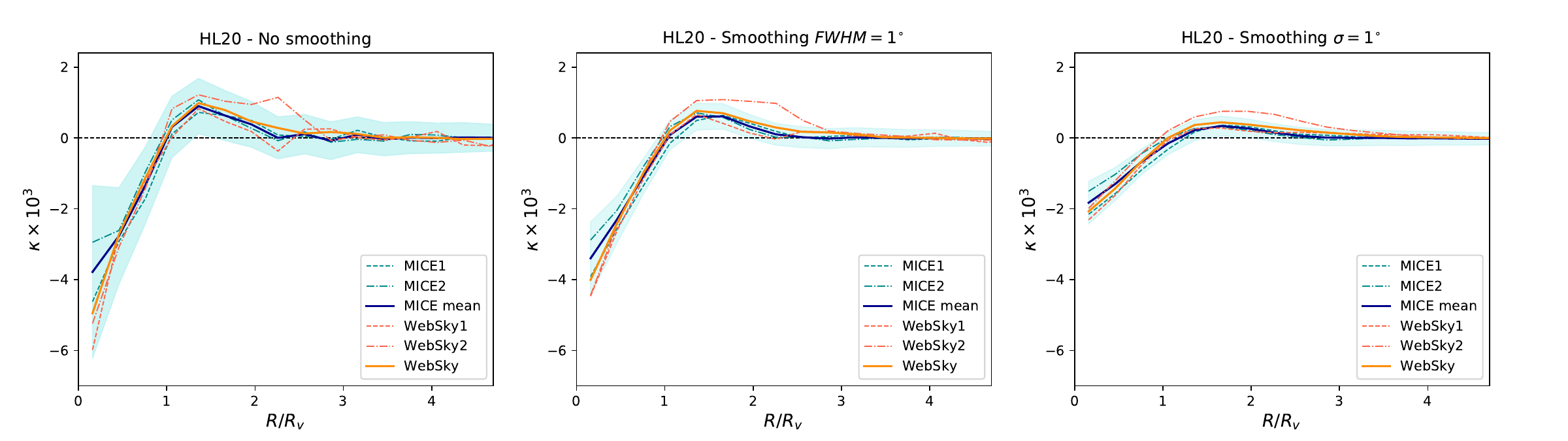}
\end{center}
\caption{Comparisons of stacked imprints of simulated voids using HL10 (top row) and HL20 (bottom row) void finder setups for the three different smoothing strategies we analyse in the paper. Dashed profiles show the the stacked imprints in different DES Y1-like patches for the MICE (blue) and WebSky (red) simulations. Solid blue lines represents our baseline estimation of the expected signal as mean of the signals from the two individual MICE patches. The solid orange profiles mark the more precise full sky estimate of the stacked signal for WebSky data. Changes due to different input cosmologies and field-to-field variations are comparable and are within the errors of our DES Y1 measurements.}
\label{fig:stack_2Dvoids_prof_websky}
\end{figure*}

\begin{table}
\centering
{\bf No smoothing}
\\
\begin{tabular}{@{}ccccc}
\hline
\hline
{\VIDE} & HD10 & HD20 & HL10 & HL20 \\
\hline
$0.91\pm0.59$ & $0.86\pm0.32$ & $1.28\pm0.38$ & $0.71\pm0.24$ & $0.93\pm0.29$ \\
\hline
\end{tabular}

{\bf \textit{FWHM}$=1^{\circ}$ smoothing}
\\
\begin{tabular}{@{}ccccc}
\hline
\hline
{\VIDE} & HD10 & HD20 & HL10 & HL20 \\
\hline
$1.23\pm0.47$ & $0.84\pm0.24$ & $1.23\pm0.32$ & $0.72\pm0.17$ & $0.85\pm0.23$ \\
\hline
\end{tabular}

{\bf $\sigma=1^{\circ}$ smoothing}
\\
\begin{tabular}{@{}ccccc}
\hline
\hline
{\VIDE} & HD10 & HD20 & HL10 & HL20 \\
\hline
$1.42\pm0.49$ & $0.83\pm0.23$ & $0.90\pm0.27$ & $1.04\pm0.22$ & $0.91\pm0.25$ \\
\hline
\end{tabular}
\caption{\label{table3} Similar to Table 2, but here amplitudes ($A$) and their errors ($\sigma_{\rm A}$) are listed for all measurement configurations for DES Y1 signals. In the case of MICE, amplitudes are all $A=1$ by definition, while the uncertainties are identical.}
\end{table}

\subsection{Testing the role of the input cosmology}
\label{websky_tests}
In Section \ref{mice_sims}, we argued that the MICE cosmological parameters with $\Omega_m=0.25$, $\sigma_8=0.8$, and $h=0.7$ may represent a sufficiently accurate description of the DES Y1 data set that we use in this study, as opposed to the best-fit {\it Planck} cosmology \citep{Planck2018_cosmo} with $\Omega_m\approx0.315\pm0.007$, $\sigma_8\approx0.811\pm0.006$, and $h=0.674\pm0.005$.

We nevertheless intended to test the shape and the amplitude of the stacked signal of voids in a simulated data set based on the {\it Planck} 2018. Therefore, we analysed the publicly available\footnote{https://mocks.cita.utoronto.ca/data/websky/} WebSky simulation \citep{websky} package. The WebSky data set provides a light-cone halo catalogue and, among other data products, a corresponding CMB lensing $\kappa$ map. An important difference is that while the MICE simulation provides realistic mock \emph{galaxy} catalogues that mimic the observed DES Y1 data, the WebSky simulation offers dark matter halo catalogues. In order to make this halo catalogue to be as DES-like as possible, we set the the same redshift range and applied a simple halo mass cut to approximately model the population of luminous red galaxies that were used as tracers of voids in our analysis. LRGs are expected to reside in halos of mass $\sim10^{13}-10^{14}h^{-1} M_\odot$ \citep[see e.g.][]{Zheng2009} that is above the mass resolution of the WebSky halo mock catalogue with $\sim10^{12}h^{-1} M_\odot$. We therefore applied a simple halo mass cut with $M>10^{13.5}h^{-1}M_\odot$ to define an LRG-like population. In particular, this selection cut is intended to model the high luminosity sample that we compare to the WebSky results below. We also added Gaussian photo-$z$ errors with a \redmagic\-like $\sigma_z/(1+z)\approx 0.02$ scatter to the simulated WebSky spec-$z$ coordinates to create realistic observational conditions.

We first identified 19,729, and 8,784 number of voids in the full-sky WebSky simulation data for our usual 10 ${\rm Mpc/h}$ and 20 ${\rm Mpc/h}$ initial Gaussian smoothing scales, respectively. We then decided to apply the same DES Y1-like mask to the full sky WebSky data set in order to test how the signal may fluctuate when measured from the full data set as opposed to smaller patches. We note that in fact we used two of these DES Y1-like survey patches in the MICE octant to estimate our signal as their mean and therefore we can also compare the field-to-field fluctuations in the MICE simulation. Therefore, this set of results facilitates a comparison of not just possible differences in the lensing imprint from changes in cosmology, but also a characterisation of simple field-to-field variations in a given cosmology, either MICE or WebSky. With the HL10 setup, we identified 839 and 874 voids in WebSky data with a DES Y1-like mask applied in two cases, and 361 and 380 voids for HL20. The number of voids is somewhat lower compared to our MICE results, but given the lack of realistic \redmagic\ galaxy mock in the case of WebSky data, such differences are not unexpected, and the results can still be compared meaningfully.

For completeness, we tested all of our different smoothing strategies applied to both the CMB $\kappa$ map and the galaxy density field given high luminosity (HL) data that promises better precision according to our MICE results and that we intend to model with our pruned WebSky halo catalogue. We present these results in Figure \ref{fig:stack_2Dvoids_prof_websky}. We found that, given our measurements errors from the DES Y1 $\times$ {\it Planck} configuration, differences in the profile from changes in input cosmology are comparable to field-to-field variations if individual DES Y1-like patches are considered either in MICE or in WebSky simulations. We therefore conclude that while in principle changes in cosmological parameters such as $\Omega_m$, $\sigma_8$, and $H_0$ may affect void lensing imprints in the CMB, our current measurements of this signal lack the precision to be sensitive to such small changes in these parameters.

\section{Discussion \& Conclusions}
\label{Section6}

The main objective of this work was to study cosmic voids identified in Dark Energy Survey galaxy samples, culled from the first year of observations. We relied on the {\redmagic} sample of luminous red galaxies of exquisite photometric redshift accuracy to robustly identify cosmic voids in photometric data. We then aimed to cross-correlate these cosmic voids with lensing maps of the Cosmic Microwave Background using a stacking methodology.

Such a signal has already been detected by \cite{cai} with a significance of $3.2\sigma$. They stacked patches of the publicly available lensing convergence map of the {\it Planck} satellite on positions of voids identified in the BOSS footprint. In general, we followed their methodology but we put more emphasis on simulation analyses to detect a signal with DES data, given different galaxy tracer density and void finding methods. In particular, we used simulated DES-like {\redmagic} galaxy catalogues together with a simulated lensing convergence map from the MICE {\it Grand Challenge} N-body simulation to test our ability to detect the CMB lensing imprint of cosmic voids.

We constrained the ratio
of the observed and expected lensing systems, which we called $A$. We first analysed the signal-to-noise corresponding to the CMB $\kappa$ profile of MICE {\redmagic} voids. We considered different void populations including 2D voids and {\VIDE} voids in 3D. We varied the galaxy density and also the initial smoothing scale applied to the density field to find the centres of the 2D voids \citep[see][for details]{Carles_void}. These parameters affect the significance of the measurement as the total number of voids, mean void size, underdensity in void interiors, and their depth in their centres are all affected by these choices and hence so is the resulting lensing signal and noise.

We then comprehensively searched for the best combination of parameters that guarantees the best chance to detect a signal with observed DES data. We concluded that the lower tracer density of the higher luminosity {\redmagic} galaxy catalogue is preferable to achieve a higher signal-to-noise for both $10$ ${\rm Mpc/h}$ and $20$ ${\rm Mpc/h}$ initial Gaussian smoothing. 

We tested to prospects of using sub-classes of voids instead of the full sample, but concluded that stacking all voids is preferable for the best measurement configuration with DES Y1 data.

We also tested the importance of post-processing in the MICE $\kappa$ map. We experimentally verified that Gaussian smoothing of scales \textit{FWHM}$=1^{\circ}$ and $\sigma=1^{\circ}$ reduce the size of the small-scale fluctuations in the lensing map while preserving most of the signal. For completeness, we created stacked images for all smoothing versions and provided a detailed comparison of the results. In the MICE analysis, we found that the best measurement configurations to detect a stacked signal are achieved when considering a 2D void catalogue with high luminosity tracers and $10$ ${\rm Mpc/h}$ initial density smoothing (HL10), exceeding $S/N\approx5$ for given $\kappa$ smoothing strategies.

We then identified voids in the observed DES Y1 {\redmagic} catalogue and compared their properties with MICE voids. In general, we found a good agreement when comparing observed 2D and {\VIDE} void catalogues with both DES Y1-like MICE mocks that we used for predictions. We repeated the simulated stacking analyses using the observed {\it Planck} CMB lensing map. The signal-to-noise is typically slightly lower than expected from MICE, due to a trend of lower amplitudes at the level of $A\approx0.8$ in some of the cases. Nevertheless, given the measurement errors, we detected a stacked signal of voids with amplitudes consistent with $A\approx1$. 

Overall, we robustly detected imprints at the $3\sigma$ significance level with most of our analysis choices, reaching $S/N\approx4$ in the best predicted measurement configurations using DES Y1 high luminosity \redmagic\ data. We found that {\VIDE} voids provided similar imprints in the CMB lensing maps, albeit at consistently lower $S/N$ than 2D voids. This finding, however, is not unexpected given the conservative cuts we apply to select our {\VIDE} sample. We leave the possible further improvements in the {\VIDE} analysis for future work.

Using the WebSky simulation, we also tested how changes in cosmological parameters might affect our results. We found that differences that arise from field-to-field variations in the signal in DES Y1-like patches, and differences due to input cosmology are comparable to each other and are within errors throughout the full imprint profile. Therefore, the level of the precision offered by a DES Y1-like data set combined with the {\it Planck} CMB $\kappa$ map is not sufficient for such precision tests. Increased galaxy survey window and a more numerous catalogue of voids, or better precision in the reconstruction of the CMB lensing fluctuations may increase the precision of these measurements in the near future.

Regarding the previously reported excess ISW signal in DES void samples compared to $\varLambda$\textit{CDM} simulations, however, we conclude that the excess in the CMB temperature maps at void locations has no counterpart in the {\it Planck} CMB lensing map. This finding does not necessarily invalidate the ISW tension. First, \cite{cai} also reported excess ISW signals using BOSS data,  but found a stacked $\kappa$ signal in good agreement with $\varLambda$\textit{CDM} simulations. Second, no detailed simulation work has jointly estimated the ISW and CMB lensing signal of voids in some alternative cosmologies. It is yet to be analysed if the excess ISW signal should always be imprinted in the corresponding CMB $\kappa$ map. Such simulation analyses could potentially exclude the coexistence of an enhanced ISW signal and a $\varLambda$\textit{CDM}-like CMB $\kappa$ imprint, pointing towards some exotic systematic effect that results in an ISW-like excess in {\it Planck} temperature data aligned with the biggest voids in both BOSS and DES data.

Our goal for the future is to create a bigger catalogue of voids, and potentially superclusters, using galaxy catalogues from three years of observed DES data (DES Y3). These presumably more accurate future detections with more voids will most probably allow cosmological parameter constraints as suggested by e.g. \cite{Chantavat2016}. Furthermore, joint analyses of CMB lensing and galaxy shear statistics may constrain modified gravity models \citep[see e.g.][]{Cautun2018, Baker2018}.

In the near future, beyond a better understanding of the methodologies, new simulations and new cosmic web decomposition data from experiments such as the Dark Energy Spectroscopic Instrument (DESI) \citep{DESI} and the Euclid mission \citep{euclid} will further constrain the lensing and ISW signals of cosmic voids.

\section*{Acknowledgments}

This work has made use of CosmoHub \citep[see][]{Carretero2017}. CosmoHub has been developed by the Port d'Informaci\'{o} Cient\'{i}fica (PIC), maintained through a collaboration of the Institut de F\'{i}sica d'Altes Energies (IFAE) and the Centro de Investigaciones Energ\'{e}ticas, Medioambientales y Tecnol\'{o}gicas (CIEMAT), and was partially funded by the ``Plan Estatal de Investigaci\'{o}n Cient\'ifica y T\'{e}cnica y de Innovaci\'{o}n'' program of the Spanish government.

Funding for the DES Projects has been provided by the U.S. Department of Energy, the U.S. National Science Foundation, the Ministry of Science and Education of Spain, the Science and Technology Facilities Council of the United Kingdom, the Higher Education Funding Council for England, the National Center for Supercomputing Applications at the University of Illinois at Urbana-Champaign, the Kavli Institute of Cosmological Physics at the University of Chicago, the Center for Cosmology and Astro-Particle Physics at the Ohio State University, the Mitchell Institute for Fundamental Physics and Astronomy at Texas A\&M University, Financiadora de Estudos e Projetos, 
Funda{\c c}{\~a}o Carlos Chagas Filho de Amparo {\`a} Pesquisa do Estado do Rio de Janeiro, Conselho Nacional de Desenvolvimento Cient{\'i}fico e Tecnol{\'o}gico and the Minist{\'e}rio da Ci{\^e}ncia, Tecnologia e Inova{\c c}{\~a}o, the Deutsche Forschungsgemeinschaft and the Collaborating Institutions in the Dark Energy Survey. 

The Collaborating Institutions are Argonne National Laboratory, the University of California at Santa Cruz, the University of Cambridge, Centro de Investigaciones Energ{\'e}ticas, Medioambientales y Tecnol{\'o}gicas-Madrid, the University of Chicago, University College London, the DES-Brazil Consortium, the University of Edinburgh, the Eidgen{\"o}ssische Technische Hochschule (ETH) Z{\"u}rich, 
Fermi National Accelerator Laboratory, the University of Illinois at Urbana-Champaign, the Institut de Ci{\`e}ncies de l'Espai (IEEC/CSIC), 
the Institut de F{\'i}sica d'Altes Energies, Lawrence Berkeley National Laboratory, the Ludwig-Maximilians Universit{\"a}t M{\"u}nchen and the associated Excellence Cluster Universe, the University of Michigan, the National Optical Astronomy Observatory, the University of Nottingham, The Ohio State University, the University of Pennsylvania, the University of Portsmouth, SLAC National Accelerator Laboratory, Stanford University, the University of Sussex, Texas A\&M University, and the OzDES Membership Consortium.

Based in part on observations at Cerro Tololo Inter-American Observatory, National Optical Astronomy Observatory, which is operated by the Association of Universities for Research in Astronomy (AURA) under a cooperative agreement with the National Science Foundation.

The DES data management system is supported by the National Science Foundation under Grant Numbers AST-1138766 and AST-1536171. The DES participants from Spanish institutions are partially supported by MINECO under grants AYA2015-71825, ESP2015-66861, FPA2015-68048, SEV-2016-0588, SEV-2016-0597, and MDM-2015-0509, 
some of which include ERDF funds from the European Union. IFAE is partially funded by the CERCA program of the Generalitat de Catalunya. 

Research leading to these results has received funding from the European Research Council under the European Union's Seventh Framework Program (FP7/2007-2013) including ERC grant agreements 240672, 291329, 306478, and 615929. We acknowledge support from the Brazilian Instituto Nacional de Ci\^enciae Tecnologia (INCT) e-Universe (CNPq grant 465376/2014-2).

This manuscript has been authored by Fermi Research Alliance, LLC under Contract No. DE-AC02-07CH11359 with the U.S. Department of Energy, Office of Science, Office of High Energy Physics.

PV acknowledges the support from the grant MIUR PRIN 2015 "Cosmology and Fundamental Physics: illuminating the Dark Universe with Euclid".

AK has been supported by a Juan de la Cierva fellowship from MINECO with project number IJC2018-037730-I. Funding for this project was also available in part through SEV-2015-0548 and AYA2017-89891-P. 

This project has also received funding from the European Union's Horizon 2020 research and innovation programme under the Marie Sk\l{}odowska-Curie grant agreement No. 754558.

\section*{Data availability}
The data and software underlying this article are available from public websites specified in various footnotes, or will be shared on reasonable request to the corresponding author.

\bibliographystyle{mnras}		  
\bibliography{biblio}

\section*{Affiliations}
$^{1}$ Institut de F\'{\i}sica d'Altes Energies (IFAE), The Barcelona Institute of Science and Technology, Campus UAB, 08193 Bellaterra (Barcelona) Spain\\
$^{2}$ SISSA, International School for Advanced Studies, Via Bonomea 265, 34136 Trieste, Italy\\
$^{3}$ IFPU, Institute for Fundamental Physics of the Universe, Via Beirut 2, 34151 Trieste, Italy\\
$^{4}$ Instituto de Astrof\'{\i}sica de Canarias (IAC), Calle Via Lactea, E-38200, La Laguna, Tenerife, Spain \\
$^{5}$ Departamento de Astrof\'{\i}sica, Universidad de La Laguna (ULL), E-38206, La Laguna, Tenerife, Spain \\
$^{6}$ Institut d'Estudis Espacials de Catalunya (IEEC), 08034 Barcelona, Spain\\
$^{7}$ Institute of Space Sciences (ICE, CSIC),  Campus UAB, Carrer de Can Magrans, s/n,  08193 Barcelona, Spain\\
$^{8}$ Department of Physics and Astronomy, University of Pennsylvania, Philadelphia, PA 19104, USA\\
$^{9}$ Universit\"ats-Sternwarte, Fakult\"at f\"ur Physik, Ludwig-Maximilians Universit\"at M\"unchen, Scheinerstr. 1, 81679 M\"unchen, Germany\\
$^{10}$ Department of Physics, University of Michigan, Ann Arbor, MI 48109, USA\\
$^{11}$ Instituci\'o Catalana de Recerca i Estudis Avan\c{c}ats, E-08010 Barcelona, Spain\\
$^{12}$ Institute of Cosmology and Gravitation, University of Portsmouth, Portsmouth, PO1 3FX, UK\\
$^{13}$ Department of Physics \& Astronomy, University College London, Gower Street, London, WC1E 6BT, UK\\
$^{14}$ Cerro Tololo Inter-American Observatory, National Optical Astronomy Observatory, Casilla 603, La Serena, Chile\\
$^{15}$ Fermi National Accelerator Laboratory, P. O. Box 500, Batavia, IL 60510, USA\\
$^{16}$ Instituto de Fisica Teorica UAM/CSIC, Universidad Autonoma de Madrid, 28049 Madrid, Spain\\
$^{17}$ Kavli Institute for Particle Astrophysics \& Cosmology, P. O. Box 2450, Stanford University, Stanford, CA 94305, USA\\
$^{18}$ SLAC National Accelerator Laboratory, Menlo Park, CA 94025, USA\\
$^{19}$ Centro de Investigaciones Energ\'eticas, Medioambientales y Tecnol\'ogicas (CIEMAT), Madrid, Spain\\
$^{20}$ Laborat\'orio Interinstitucional de e-Astronomia - LIneA, Rua Gal. Jos\'e Cristino 77, Rio de Janeiro, RJ - 20921-400, Brazil\\
$^{21}$ Department of Astronomy, University of Illinois at Urbana-Champaign, 1002 W. Green Street, Urbana, IL 61801, USA\\
$^{22}$ National Center for Supercomputing Applications, 1205 West Clark St., Urbana, IL 61801, USA\\
$^{23}$ Physics Department, 2320 Chamberlin Hall, University of Wisconsin-Madison, 1150 University Avenue Madison, WI  53706-1390\\
$^{24}$ INAF-Osservatorio Astronomico di Trieste, via G. B. Tiepolo 11, I-34143 Trieste, Italy\\
$^{25}$ Institute for Fundamental Physics of the Universe, Via Beirut 2, 34014 Trieste, Italy\\
$^{26}$ Observat\'orio Nacional, Rua Gal. Jos\'e Cristino 77, Rio de Janeiro, RJ - 20921-400, Brazil\\
$^{27}$ Department of Physics, IIT Hyderabad, Kandi, Telangana 502285, India\\
$^{28}$ Department of Astronomy/Steward Observatory, University of Arizona, 933 North Cherry Avenue, Tucson, AZ 85721-0065, USA\\
$^{29}$ Jet Propulsion Laboratory, California Institute of Technology, 4800 Oak Grove Dr., Pasadena, CA 91109, USA\\
$^{30}$ Santa Cruz Institute for Particle Physics, Santa Cruz, CA 95064, USA\\
$^{31}$ Kavli Institute for Cosmological Physics, University of Chicago, Chicago, IL 60637, USA\\
$^{32}$ Department of Astronomy, University of Michigan, Ann Arbor, MI 48109, USA\\
$^{33}$ Department of Physics, Stanford University, 382 Via Pueblo Mall, Stanford, CA 94305, USA\\
$^{34}$ Department of Physics, ETH Zurich, Wolfgang-Pauli-Strasse 16, CH-8093 Zurich, Switzerland\\
$^{35}$ Center for Cosmology and Astro-Particle Physics, The Ohio State University, Columbus, OH 43210, USA\\
$^{36}$ Department of Physics, The Ohio State University, Columbus, OH 43210, USA\\
$^{37}$ Center for Astrophysics $\vert$ Harvard \& Smithsonian, 60 Garden Street, Cambridge, MA 02138, USA\\
$^{38}$ Australian Astronomical Optics, Macquarie University, North Ryde, NSW 2113, Australia\\
$^{39}$ Lowell Observatory, 1400 Mars Hill Rd, Flagstaff, AZ 86001, USA\\
$^{40}$ Departamento de F\'isica Matem\'atica, Instituto de F\'isica, Universidade de S\~ao Paulo, CP 66318, S\~ao Paulo, SP, 05314-970, Brazil\\
$^{41}$ George P. and Cynthia Woods Mitchell Institute for Fundamental Physics and Astronomy, and Department of Physics and Astronomy, Texas A\&M University, College Station, TX 77843,  USA\\
$^{42}$ Department of Astrophysical Sciences, Princeton University, Peyton Hall, Princeton, NJ 08544, USA\\
$^{43}$ School of Physics and Astronomy, University of Southampton,  Southampton, SO17 1BJ, UK\\
$^{44}$ Computer Science and Mathematics Division, Oak Ridge National Laboratory, Oak Ridge, TN 37831\\
$^{45}$ Excellence Cluster Origins, Boltzmannstr.\ 2, 85748 Garching, Germany\\
$^{46}$ Max Planck Institute for Extraterrestrial Physics, Giessenbachstrasse, 85748 Garching, Germany\\
$^{47}$ Institute for Astronomy, University of Edinburgh, Edinburgh EH9 3HJ, UK\\

\end{document}